\documentclass[12pt]{article}
\usepackage{amssymb}

\usepackage{natbib,graphicx,setspace,lscape,longtable}
\usepackage{amsmath,amsthm,bm}
\usepackage{fullpage}  

\usepackage[usenames,dvipsnames]{color}
\usepackage{graphicx}               

\definecolor{DarkGreen}{rgb}{0.5,0.8,0.6}   
\definecolor{RGBblack}{rgb}{0.0,0.0,0.0}    

\def\stackover#1#2{\mathrel{\mathop{#2}\limits^{#1}}}
\newcommand{\iid}{\stackover{\mbox{\footnotesize i.i.d.}}{\sim}}

\newcommand{\is}{\itemsep=0pt}
\newcommand{\bd}[1]{\begin{description}[#1]\is}
  \newcommand{\ed}{\end{description}}
\newcommand{\bi}{\begin{itemize}\is}
  \newcommand{\ei}{\end{itemize}}
\newcommand{\be}{\begin{enumerate}\is}
  \newcommand{\ee}{\end{enumerate}}
  \newcommand{\beq}{\begin{eqnarray}\is}
  \newcommand{\eeq}{\end{eqnarray}}
\makeatletter
\newcommand*{\rom}[1]{\expandafter\@slowromancap\romannumeral #1@}

\newcommand{\bZ}{\bm Z}
\newcommand{\bx}{\bm x}

\newcommand{\blambda}{\bm \lambda}

\newcommand{\bbeta}{\bm \beta}

\newcommand{\HH}{\mathcal{H}}
\newcommand{\Sbar}{\overline{S}}

\renewcommand{\th}{\theta}

\newcommand{\sig}{\sigma}
\newcommand{\Sig}{\Sigma}
\newcommand{\LN}{\mbox{LN}}
\newcommand{\N}{\mbox{N}}
\newcommand{\Be}{\mbox{Be}}
\newcommand{\Unif}{\mbox{Unif}}
\newcommand{\Ga}{\mbox{Ga}}

\newcommand{\phat}{\widehat{p}}
\newcommand{\DDPGP}{\mbox{DDP-GP}}
\newcommand{\IPTW}{\mathrm{IPTW}}

\newcommand{\nstate}{n_{\mbox{state}}}
\newcommand{\nstage}{n_{\mbox{stage}}}
\newcommand{\ntrans}{n_{T}}

\begin{document}
\doublespacing

\title{Bayesian Nonparametric Estimation for Dynamic Treatment Regimes
with Sequential Transition Times} 
\author{
            Yanxun Xu\\
            {\small Division of Statistics and Scientific Computing, The University of Texas at Austin, Austin, TX}
            \and Peter M\"uller \thanks{Address for Correspondence:  Department of Mathematics
UT Austin 1, University Station, C1200, Austin, TX 78712 USA. E-mail: pmueller@math.utexas.edu.}\\
            {\small Department of Mathematics, The University of Texas at Austin, Austin, TX} 
             \and Abdus S. Wahed\\
          {\small   Department of Biostatistics, University of Pittsburgh, Pittsburgh, PA}   
             \and Peter F. Thall\\
        {\small Department of Biostatistics, The University of Texas M.D. Anderson Cancer Center, Houston, TX}
}           
\date{}
\maketitle

\begin{abstract}
Dynamic treatment regimes in oncology and other disease areas
often can be characterized by an alternating sequence of
treatments or other actions and transition times between disease states.
The sequence of transition states may vary substantially from patient to patient,
depending on how the regime plays out, and in practice there often are  many possible
counterfactual outcome sequences. For evaluating the regimes, the mean final overall time
 may be expressed as a weighted average of the means of all possible sums of successive transitions times.
A common example arises in cancer therapies where the transition times between
various sequences of treatments, disease remission, disease progression, and death
characterize overall survival time. For the general setting,
we propose estimating mean overall outcome time  by assuming a Bayesian nonparametric
regression model for the logarithm of each transition time.
A dependent Dirichlet process prior with
Gaussian process base measure (DDP-GP) is assumed, and
a joint posterior is obtained by Markov chain Monte Carlo (MCMC) sampling.
We provide general guidelines for constructing a prior using empirical Bayes methods. We compare the proposed approach with  inverse probability of treatment weighting.
These comparisons are done by simulation studies of both single-stage
and multi-stage regimes,
with treatment assignment depending on baseline covariates.  The method is
applied to analyze a dataset arising from a clinical trial involving multi-stage chemotherapy regimes for acute leukemia.
An R program for implementing the DDP-GP-based Bayesian nonparametric analysis is freely available at
https://www.ma.utexas.edu/users/yxu/.

\noindent{\bf KEY WORDS:}  Dependent Dirichlet  process;  Gaussian process; G-Estimation; Inverse probability of treatment weighting; Markov chain Monte Carlo.
\end{abstract}

\section{Introduction}
\label{sec:intro}

We propose a Bayesian nonparametric (BNP) approach for evaluating
dynamic treatment regimes (DTRs) in which the outcome at each stage is
a random transition time between two disease states.  The final
outcome of primary interest is the sum, $T$, of a sequence of
transition times. The sequence of transition times  that are actually observed
is 
determined by the way that the patient's treatment regime plays out.
The mean of $T$ may be expressed as an appropriately weighted average
over all possible sequences of event times. 
For example, with fatal diseases $T$ often is overall
survival (OS) time. An algorithm commonly used by oncologists in
chemotherapy of solid tumors is to choose the patient's initial
(frontline) treatment based on his/her baseline covariates, continue as long as the patient's disease is stable, switch to
a different chemotherapy (salvage) if progressive disease ($P$)
occurs, stop chemotherapy if the tumor is brought into complete or
partial remission ($C$), and begin salvage if $P$ occurs at some time
after $C.$ There are many elaborations of this in oncology, including
multiple attempts at salvage therapy, use of consolidation therapy for
patients in remission, suspension of therapy if severe toxicity is
observed, or inclusion of radiation therapy or surgery in the regime.
An important potential application of this structure is treatment
regimes for psychological disorders or drug addiction.  For example,
in treatment of schizophrenia one may replace $P$ by a psychotic
episode or other worsening of the subject's psychological status, $C$
by a specified improvement in mental status, and death by a
psychological breakdown severe enough to require hospitalization.

\begin{figure}
\centering
\includegraphics[width=.6\textwidth]{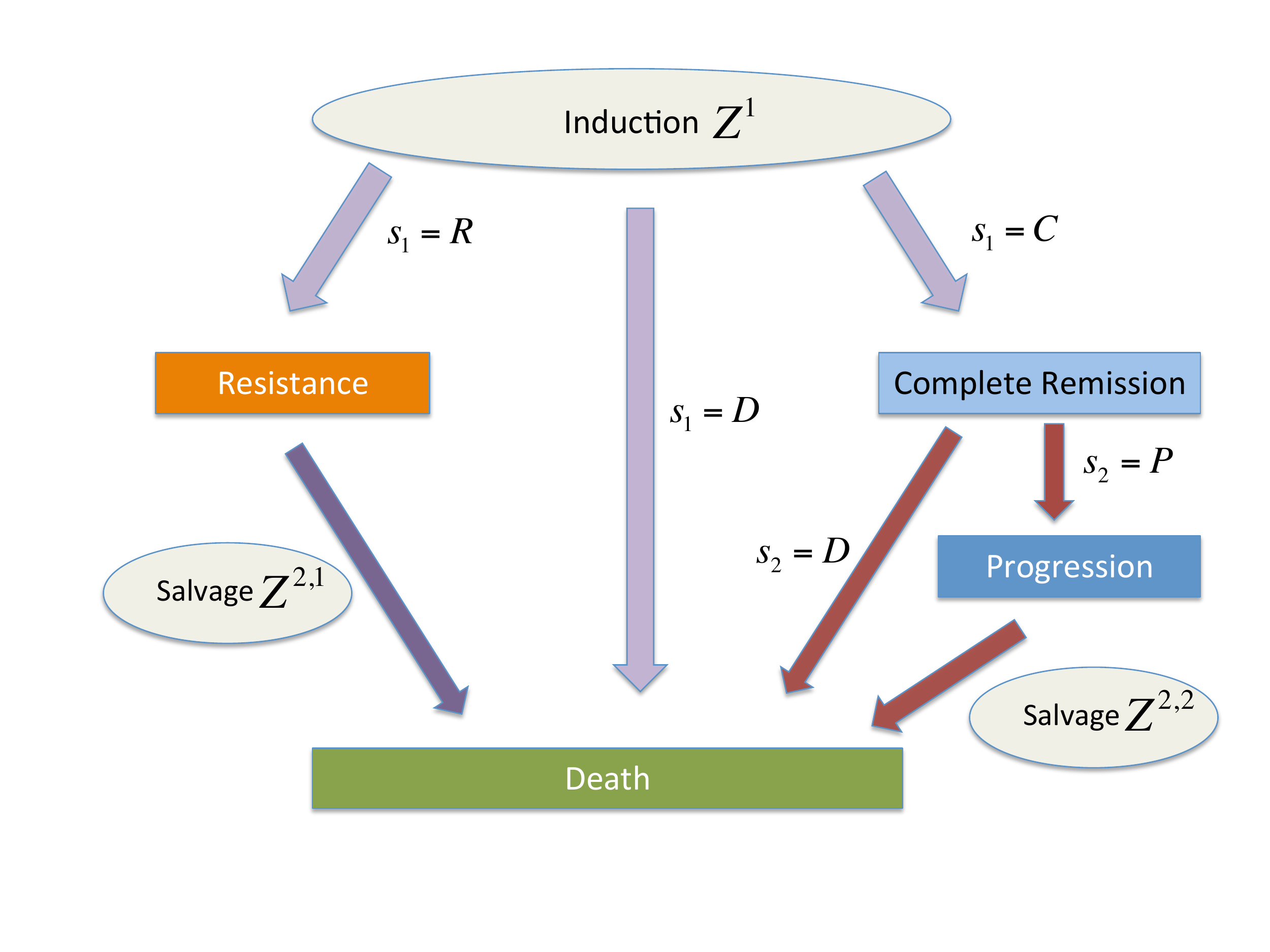}
\caption{The scheme }
\label{fig:flowchart}
\end{figure}

Denote the action at stage $\ell$ of the DTR by $Z^\ell$,
which may be a treatment or a decision to delay or terminate therapy.
Here, the term stage refers to the decision points in the DTR --
that is, the choice of frontline and possible salvage therapies.
At each stage we observe some disease state $s_\ell$, such as 
$P, C$ or death ($D$). 
Let $T^{j,r}$ denote the transition time from disease state $j$ to state
$r$, with $j=0$ the patient's initial disease status.  
See Figure \ref{fig:flowchart} for an example
(details of which will be provided later)
with up to $\nstage=3$ stages, $\nstate=4$ disease states, and a total
of $\ntrans = 7$ possible 
transition times. The actual number of stages and observed 
transition times varies across patients and depends on the specific
treatment-outcome sequence.  A DTR is the sequence ${\bf Z}$ = $(Z^1, Z^2,
\cdots)$, where each $Z^\ell$ is an adaptive action based on the
patient's history $\HH^{\ell-1}$ of previous treatments and
transition times, and 
$\HH^0$ refers to baseline covariates.
One possible treatment-outcome sequence is 
$(\HH^0, Z^1, T^{0,P}, Z^2, T^{P,D}),$ in which the initial chemotherapy
$Z^1$ was chosen based on $\HH^0$, $Z^2$ was chosen based on
$\HH^1$ = $(\HH^0, Z^1, T^{0,P})$ and given at time
$T^{0,P}$ of $P$, and OS time $T$ = $T^{0,P} + T^{P,D}.$ Similarly, a
patient brought into remission who later suffers progressive disease
has sequence $(\HH^0, Z^1, T^{0,C}, T^{C,P}, Z^2, T^{P,D})$ and
$T$ = $T^{0,C} + T^{C,P} + T^{P, D}.$
We will focus on application of BNP methods for estimating the
conditional distributions of the transition times give the most recent
histories, with the goal to estimate the mean of $T$ for each possible
DTR.  Key elements of our proposed approach are quantification of all
sources of uncertainty and prediction of $T$ under a reasonable set of
viable counterfactual DTRs \citep{wang2012evaluation}.  BNP methods have
been used in estimating regime effects by 
\cite{hill2011bayesian} and 
\cite{karabatsos2012bayesian}.

Since all elements of a DTR may affect $T$, the clinically relevant
problem is optimizing the entire regime, rather than the treatment at
one particular stage. Most clinical trials or data analyses attempt to
reduce variability by focusing on one stage of the actual DTR, usually
frontline or first salvage treatment, or by combining stages in some
manner.  This often misrepresents actual clinical practice, and
consequently conclusions may be very misleading.  For example, an
aggressive frontline cancer chemotherapy may maximize the probability
of $C$, but it may cause so much immunologic damage that any salvage
treatment given after rapid relapse, i.e. short $T^{C,P}$, may be
unlikely to achieve a second remission.  In contrast, a milder
induction treatment may be suboptimal to eradicate the tumor, but it
may debulk the tumor sufficiently to facilitate surgical
resection. Such synergies may have profound implications for clinical
practice, especially because effects of multi-stage treatment regimes
often are not obvious and may seem counter-intuitive.  Physicians who
have not been provided with an evaluation of the composite effects of
entire regimes on the final outcome may unknowingly set patients on
pathways that include only inferior regimes.

A major practical advantage of BNP models is that they often provide
better fits to complicated data structures than can be obtained using
parametric model-based methods.  In our motivating application,
where leukemia patients were randomized among initial chemotherapy
treatments but not among later salvage therapies, the BNP model
provides good fits for each transition time distribution conditional
on previous history.  Failure to randomize patients in treatment
stages after the first is typical in clinical trials, most of which
ignore all but the first stage of therapy.  In contrast, sequential
multi-arm randomized treatment (SMART) designs, wherein patients are
re-randomized at stages after the first, have been used in oncology
trials \citep{thall2000evaluating, thall2007adaptive, thall2007bayesian}, and are being used increasingly
in trials to study multi-stage adaptive regimes for behavioral or
psychological disorders \citep{dawson2004placebo, murphy2007customizing, murphy2007developing, connolly2007practice}.

While re-randomization is desirable, it is not commonly done and
inference has to adjust for this lack of randomization. 
A wide array of methods have been proposed for evaluating DTRs from
observational data and longitudinal studies, beginning with the
seminal papers by \cite{robins1986new,robins1987addendum,robins1989analysis,robins1997causal} on G-estimation of
structural nested models.  Additional references include applications
to longitudinal data in AIDS \citep{hernan2000marginal}, 
inverse probability of treatment weighted (IPTW) estimation of
marginal structural models \citep{murphy2001marginal, van2007causal, robins2008estimation}, G-estimation for optimal DTRs \citep{murphy2003optimal, robins2004optimal}, and a review by \cite{moodie2007demystifying}. A variety of methods have been developed to evaluate DTRs from clinical trials \citep{lavori2000design,thall2002selecting,murphy2005experimental}. For survival analysis, \cite{lunceford2002estimation} introduced {\it ad hoc}
estimators for the survival distribution and mean restricted survival
time under different treatment policies. These estimators, although
consistent, were inefficient and did not exploit information from auxiliary covariates.
\cite{wahed2006semiparametric} derived
more efficient, easy-to-compute estimators that included auxiliary covariates
for the survival distribution and related quantities of DTRs.
Their estimators compared DTRs using data
from a two-stage randomized trial, in which two options were available
for both stages and the second-stage treatment assignments were determined
by randomization. However, these estimators must be adapted for
more general or more complicated designs that permit various numbers of
treatment options at each stage and involve the scenarios where
second-stage treatment is not randomized, but rather
determined by the attending physicians.

In settings where the DTR's final overall time, such as survival time,
is the sum of a sequence of transition times, we propose a Bayesian
nonparametric approach that employs a 
nonparametric regression model for (the
logarithm of) each transition time conditional on the most recent
history of actions and outcomes.  We assume a dependent Dirichlet
process prior with Gaussian process base measure (DDP-GP), and compute
 a joint posterior by Markov chain Monte Carlo (MCMC) sampling.  To
address the important issue that Bayesian analyses depend on prior
assumptions, we provide guidelines for using empirical Bayes methods
to establish prior hyperparameters.  Posterior analyses include
estimation of posterior mean overall outcome times and credible
intervals for each DTR.

The rest of the paper is organized as follows. In Section 2
we review the motivating study, and give a brief review of DTRs in
settings with successive transition times in Section 3.  We present
the DDP-GP model in Section 4. A simulation study of the BNP approach
in single-stage and multi-stage regimes, with comparison to
frequentist IPTW, is summarized in Section 5.  We re-analyze the
leukemia trial data in Section 6, and close with brief
discussion in Section 7.

\section{Motivating Study}
 Our proposed methodology is motivated by a clinical trial conducted at The University of Texas M.D. Anderson Cancer Center to evaluate chemotherapies for
acute myelogenous leukemia (AML) or
myelo-dysplastic syndrome (MDS).
Patients were  randomized fairly among  four front-line combination chemotherapies for remission induction: fludarabine + cytosine arabinoside  (ara-C)
plus idarubicin (FAI), FAI + all-trans-retinoic acid (ATRA), FAI
+ granulocyte colony stimulating factor (GCSF), and FAI + ATRA
+ GCSF.  The goal of  induction
therapy for AML/MDS is to achieve complete remission ($C$),  a necessary
but not sufficient condition for long-term survival.
Patients who do not achieve $C$, or who achieve $C$ but
later relapse, are given salvage
treatments as another attempt to achieve $C$.
Following conventional clinical practice, patients were not  randomized among salvage therapies,
which were chosen by the attending physicians based on clinical judgment.
Since there were many types of salvage, these are broadly classified into two categories as either containing high dose ara-C (HDAC) or not.
This data set was analyzed initially using
conventional methods \citep{estey1999randomized}, including logistic regression, Kaplan-Meier
estimates, and Cox model regression, including
comparisons of the induction therapies in terms of OS that ignored possible effects of  salvage therapies.

Figure \ref{fig:flowchart} illustrates the actual possible  therapeutic pathways and
outcomes of the patients during the trial, which is typical of chemotherapy for AML/MDS.
Death might occur (1) during induction therapy, (2) following
salvage therapy if the disease was resistant to induction, (3)
during $C$, or (4) following disease progression after $C$.
\cite{wahed2013evaluating} re-analyzed
the data from  this trial  by accounting for the
structure in Figure \ref{fig:flowchart}, and identified 16 DTRs
including both frontline and salvage therapies.
To correct for bias due to the lack of randomization in estimating the mean OS times,
they  used both IPTW \citep{robins1992recovery} and G-estimation based on a frequentist likelihood.
In the G-estimation,  for each transition time they  first fit
accelerated failure time (AFT) regression models 
using Weibull, exponential, log-logistic or log-normal
distributions, and chose the distribution having smallest Bayes information
criterion (BIC).
They then performed likelihood-based G-estimation by first fitting each conditional transition time distribution regressed on patient baseline covariates and previous transition times, and then averaging over the empirical covariate distribution.

Like Wahed and Thall,  the primary goal of our analyses of the AML/MDS dataset is to estimate mean OS and determine the optimal regime. We build on their approach by replacing the parametric AFT models for transition times with the DDP-GP model.
We also demonstrate the usefulness  of the BNP regression model for G-estimation in simulation
studies of single-stage and multi-stage regimes in which treatment assignments depend on patient covariates.

\section{Dynamic Regimes with Stochastic Transition Times}

Denote the set of possible disease states
by $\{0, 1,\cdots, \nstate \},$ 
with $0$ denoting the patient's initial state before receiving the first
treatment.  The pairs of states $(s_{\ell-1},s_{\ell})$ for which a
transition $s_{\ell-1} \rightarrow s_{\ell}$ is possible at stage
$\ell$ of the patient's therapy depend on the particular regime. Here $s_0$ refers to the patient's initial state, before start of therapy. 
We will identify specific states using letters
such as $P$, $C$, etc., as in the earlier examples,  to replace the
generic integers. For example, in cancer therapy, $s_{\ell-1} \rightarrow
C$ means that a patient's disease has responded to treatment,
$P \rightarrow D$  means a patient with progressive disease has died,
and of course  $D \rightarrow s_\ell$ is impossible. We denote the
transition time from state $s_{\ell-1}$ to state $s_\ell$ in stage $\ell$ of
treatment by $T^{\ell,(s_{\ell-1},s_\ell)}$,
for ${\ell} = 1, \cdots, \nstage,$ the  maximum number of stages in the DTR. When no ambiguity arises we simply write $T^{r,s}$ for the transition time from state $r$ to $s$. 
To simplify notation for the transition time distributions, we denote
the history of all covariates, treatments, and previous transition
times through $\ell$ stages, before observation
of $T^{\ell,(s_{\ell-1},s_\ell)}$
but including the stage $\ell$ action $Z^{\ell}$ by ${\bx}^\ell$ = $({\cal
H}^{\ell-1},Z^\ell)$ = $({\bx}^0, Z^1, T^{1,(s_0,s_1)}, \cdots, Z^\ell)$, with
${\bx}^0$ = $\HH^0$.  Thus, a DTR is 
${\pmb Z}$ = $(Z^1,Z^2,\ldots)$, a sequence of  actions for all
possible stages.  When no meaning is lost, we will write 
$T^{\ell,(s_{\ell-1},s_\ell)}$ as $T^k,$
where $k=1,\ldots,\ntrans$ is a running index of all possible state
transitions. For example, in Figure \ref{fig:flowchart} we have
 up to $\nstage=3$ stages and $\ntrans=7$ possible transitions.
Similarly, we will write $\bx^k$ for the corresponding covariate vector.
 Our use of a single
index to identify stage is a slight abuse of notation since, for
example, the actual second stage of therapy might differ depending on
the sequence of outcomes. For example, stage 2 treatment $Z^2$ of a
patient with sequence $({\bx}^0, Z^1, T^{1,(0,R)}, Z^2)$ is first
salvage for resistant disease during induction with $Z^1$, while
stage 3 treatment $Z^3$ of a patient with sequence $({\bx}^0, Z^1,
T^{1,(0,C)}, T^{2,(C,P)}, Z^3)$ is first salvage for progressive
disease after achieving response initially with $Z^1.$ This latter
example could be elaborated if, under a different regime,  consolidation therapy, $Z^2$, were
given for patients who enter $C$, in which case the sequence would be
$({\bx}^0, Z^1, T^{1,(0,C)}, Z^2, T^{2,(C,P)}, Z^3).$

Below we will develop 
a general BNP model for all possible conditional
distributions of the form  $p(T^{k}~|~{\bx}^{k})$ =
$p(T^{\ell,(s_{\ell-1},s_\ell)}~|~{\bx}^{\ell}).$ This determines
the likelihood for all possible sequences of treatments and transition
times through $\ntrans$ transitions as the product
\begin{equation}
{\cal L} \ =\ \prod_{k=1}^{\ntrans}\,p(T^{k}~|~{\bx}^{k}).
\end{equation}
The overall time for any counterfactual sequence of transition times is the sum
$T\ =\ \sum_{k=1}^{\ntrans}\,T^k$.
Our goal is to
estimate the mean of $T$ for each possible ${\pmb Z}.$

\section{ A Nonparametric Bayesian Model for DTR's}
\label{sec:BNP}
\subsection{DDP and Gaussian Process Prior}
\label{sec:introDDP}

To specify the BNP model, we denote $Y^k$ = log$(T^k)$ and write the
distribution of $[Y^k~|~{\bx}^k]$ as $F^k(\cdot~|~ {\bx}^k)$.  For
convenience, we will refer to ${\bx}^k$ as \lq covariates'.  We
construct a BNP survival regression model for each $F^k(\cdot~|~
{\bx}^k)$ by successive elaborations, starting with a model for a
discrete random distribution $G(\cdot)$.
We then use a Gaussian kernel to
extend this to a prior for a continuous random distribution
$F(\cdot)$,  and
finally endow the kernel means with a regression structure by
expressing them as functions of ${\bx}^k.$
 The latter construction extends $F$ to a family $\{F(\cdot \mid
\bx^k)\}$, indexed by $\bx^k$. 
The construction of $G(\cdot)$ and $F(\cdot)$ is briefly outlined
below, by way of a brief review of BNP models. In the end we will only
use the last model $\{F(\cdot \mid \bx^k)\}$, which we use as sampling model for $Y^k$.
See, for example,  \cite{muller2013bayesian}
and  \cite{muller2013nonparametric} for
more extensive reviews of BNP inference.

The Dirichlet process (DP) prior first was proposed by 
\cite{ferguson1973bayesian} as a probability distribution on a
measurable space of probability measures.  The DP is indexed by two
hyperparameters, a base measure, $G_0,$ and a precision parameter,
$\alpha > 0$.  If a random distribution $G$ follows a DP prior, we
denote this by $G\sim DP(\alpha, G_0)$.  Denoting a beta distribution
by $\Be(a,b)$, if $G\sim DP(\alpha, G_0)$ then $G(A) \sim \Be\{ \alpha
G_0(A), \alpha[1-G_0(A)]\}$ for any measurable set $A$, and in
particular $E\{G(A)\}=G_0(A).$ Let $\delta(\theta)$ denote a point
mass at $\theta$.  \cite{sethuraman1991constructive}
provided a useful representation of the DP as
$G=\sum_{h=0}^{\infty}w_h\delta(\theta_h)$, where
$\theta_h \iid G_0$,
and the weights $w_h$ are generated sequentially from rescaled Beta
distributions as
 $w_h/(1-\sum_{r=1}^{h-1}w_r) \sim Be(1,\alpha)$,
the so-called \lq\lq stick-breaking" construction.
The discrete nature of $G$ is awkward in many applications.
A DP mixture model extends the DP model by replacing each point mass
$\delta(\theta_h)$ with a continuous kernel centered at $\theta_h$.
Without loss of generality, we will use a normal kernel.  Let
$N(\cdot;\; \mu,\sig)$ denote the measure of a normal distribution
with mean $\mu$ and standard deviation (sd) $\sig$.  The DP mixture
model assumes
\begin{equation}
  G = \sum_{h=0}^{\infty}\ w_h N(\cdot\;; \theta_h,\sig). 
\label{eq:dpm}
\end{equation}
The use and interpretation of \eqref{eq:dpm} is very similar to that
of a finite mixture of normal models. 
In practical applications, the
sum in \eqref{eq:dpm} is  often truncated at a reasonable
finite value. This 
model is useful for density estimation under i.i.d. sampling from an
unknown distribution, and it provides good fits to a wide variety of
datasets because a mixture of normals can closely approximate
virtually any distribution \citep{ishwaran2001gibbs}.

To include the regression on covariates that we will need for the
survival model of each conditional transition time distribution,
$(Y^k~|~{\bx}^k)$, we extend the DP  mixture to 
a dependent DP (DDP), which was first proposed by
\cite{maceachern1999dependent}.  The basic idea of a DDP is
 to endow each $\theta_h^k$ with additional structure that specifies how it varies as a function of covariates
 ${\bx}^k.$  Writing this regression function as  $\theta^k_h({\bx}^k)$
 for each summand in $(\ref{eq:dpm})$,
 and returning to the conditional transition time distributions, we assume that
\begin{equation}
  F^k(y ~|~{\bx}^k) = \sum_{h=0}^{\infty} w^k_h\, N(y;\ \th_h^k({\bx}^k),\,\sigma^k).
\label{eq:Fx}
\end{equation}
 This form of the DDP, which includes both the convolution with a normal kernel and functional dependence on covariates, provides a very flexible regression model.

To complete our specification of the DDP, we will assume that the $\theta^k_h(\cdot)$'s are
independent realizations from a Gaussian process prior.
The Gaussian process first was popularized by \cite{Hagan1978} in
Bayesian inference for a random function (unrelated to the use in a
DDP prior).
For more recent discussions see, for example, \cite{Rasmussen:2006, Neal1995,Shi2007}.  Temporarily suppressing the
transition index $k$ and running index $h$
in $(\ref{eq:Fx})$,
and denoting ${\bx}$ = $(x_1,\cdots,x_n)$,  a Gaussian process is a
stochastic process $\theta(\cdot)$
in which $\theta({\bx})$ = $\{\th(x_1), \ldots, \th(x_n)\}^\prime$ has a multivariate normal distribution
with mean vector $\mu({\bx})=(\mu(x_1),\ldots,\mu(x_n))$ and  $(n\times n)$ covariance matrix $C({\bx})$
for  $\bx$ of any dimension $n\geq 1$.  We denote this by $\theta \sim  GP(\mu, C).$

We use the GP prior to define the dependence of $\th_h^k(\bx^k)$
as a function of $\bx^k$ by assuming  
$\{\theta_h^k(\bx_k)\} \sim \ GP(\mu_h^k, C^k)$, 
as a function of $\bx_k$, for fixed $h$.  We will refer to
the DDP with a convolution using a normal kernel and a Gaussian
process prior on the normal kernel means as a DDP-GP model. 
While the mean and covariance processes of the GP can be quite general,
in practice,  $C^k({\bx}^k)$ often is parameterized as a
function $C^{k}_{ij} = C(x^k_i, x^k_j; \xi^k)$, where $\xi^k$ is a
vector of hyperparameters, and the mean function
is indexed similarly by hyperparameters $\bbeta_h^k$ and written as $\mu_h^k(\bx^k \mid \bbeta_h^k)$.
In the DTR setting, since each covariate vector ${\bx}^k$ is a history,
its entries can include baseline covariates,  transition times,
and indicators of previous treatments or actions.
To obtain numerically reasonable parameterizations of the Gaussian
process functions $C^k$ and $\mu_h^k$, we standardize numerical-valued
covariates such as age. 
We now have
$$
  \{\theta^k_h(\bx^k)\} \sim \ GP(\mu_h^k(\bx^k), C^k(\bx^k)), \ \ \ \ \ \ h=1, 2,...
$$
To specify the form of  $\mu_h^k$ and $C^k,$
let $i=1,2, \cdots,$  index patients,  so that  $\bx^k_{i}$ is
the history of patient $i$ at transition $k$, and define  the  indicator
$\delta_{ij}=I(i=j)$ = 1 if $i=j$ and 0 otherwise.
We model the mean function $\mu_h^k(\cdot)$ as a linear regression, by
assuming that 
\begin{equation}
   \mu_h^{k}(\bx^k_{i}) = \bx^k_{i}\bbeta_h^{k}.
\label{mux}
\end{equation}
 For patients $i$ and $j$, we assume that the covariance process takes the form
\beq
  C^k(\bx^k_{i}, \bx^k_{j}) = \exp\{ -\sum_{m=1}^{M^k}
  (x^k_{im}-x^k_{jm})^2  \} + 
  \delta_{ij}J^2,  \ \ \ i, j=1, \dots, n,
\label{eq:cov}
\eeq
where $M^k$ 
is the number of covariates
at transition $k$ and $J$ is the variance on the diagonal reflecting
the amount of jitter \citep{Neal1998}, which usually takes a small
value (e.g, $J=0.1$).
For binary covariates, the 
quadratic form in \eqref{eq:cov} 
reduces to counting the  number of binary covariates in which two
patients differ.
If desired, additional hyperparameters could be introduced in
(\ref{eq:cov}) to obtain more flexible covariance functions.
However, in practice this form of the covariance matrix
yields a strong correlation for observations 
on patients with very similar $\bx^k$, 
and has been  adopted widely  \citep{Williams1998}.

Combining all of these structures, we denote the model for conditional
distribution of the $k^{th}$ transition time as
\begin{equation}
  F^k \sim \DDPGP\left\{\{\mu_h^k\}, C^k; \alpha^k,\,\{\bbeta_h^k\},\,\sig^k\right\},
\end{equation}
recalling that the weights of the DDP are generated sequentially as
 $w^k_h/(1-\sum_{r=1}^{h-1}w^k_r) \sim Be(1,\alpha^k)$.
For later reference we state the full model,
\begin{gather}
  p(y_i^k \mid \bx_i^k, F^k) = F^k(y_i^k \mid \bx_i^k)\nonumber \\
  F^k \sim \DDPGP\left\{\{\mu_h^k\}, C^k;
    \alpha^k,\,\{\bbeta_h^k\},\,\sig^k\right\}
  \label{fullModel}
\end{gather}
$k=1,\ldots,\ntrans$.

\subsection{Determining Prior Hyperparameters}
As priors for $\bbeta_h^k$ in \eqref{mux} 
we assume $\bbeta_h^{k} \sim \ N(\bbeta_0^{k},\Sig_0^{k})$ for each
transition $k$, $h=1, 2, \dots$.  For the normal pdfs of the DDP mixture models we
assume the precision parameters follow the same prior $(\sig^k)^{-2}
\iid\ \Ga(\lambda_1,\lambda_2)$ and, similarly, for the parameters
that determine the weights of the DDP mixture under the stick-breaking
construction we assume $\alpha^k  \iid\ \Ga(\lambda_3,\lambda_4).$

To apply the DDP-GP model, one must first determine numerical values
for the fixed hyperparameters $\{\bbeta_0^k,$ $\Sig_0^k, k=1,2,... \}$
and ${\blambda}$ = $(\lambda_1,\lambda_2,\lambda_3,\lambda_4).$ This
is a critical step. These numerical hyperparameter values must
facilitate posterior computation, and they should not 
introduce inappropriate information into the prior that would
invalidate posterior inferences.
With this in mind, the hyperparameters $(\bbeta_0^k,\Sig_0^k)$ for the
$k^{th}$ transition time covariate effect distribution may be obtained
via empirical Bayes by doing a preliminary fits of a lognormal
distribution $Y^k$ = log$(T^k) \sim$ $N(\bx^k\bbeta^k_0, \sigma^k_0)$
for each transition $k.$ The covariate effect estimates then can be
used as the values of $\bbeta^k_0$.  Similarly, we assume a
diagonal matrix for $\Sig_0^k$ with the diagonal values also obtained
from the preliminary fit of the lognormal distribution.  Once an
empirical estimate of $\sigma^k$ is obtained, one can tune
$(\lambda_1, \lambda_2)$ so that the prior mean of $\sigma^k$ equals
the empirical estimate and the variance equals 1 or a suitably large
value to ensure a vague prior.  In contrast, information
about $\alpha^k$ typically is not available in practice, and an
empirical Bayes approach cannot be applied to determine $(\lambda_3,
\lambda_4)$.  However, setting $\lambda_3 = \lambda_4$ = 1 gives a
Gamma(1, 1) distribution, which has mean 1 and variance 1, and is a
well behaved, noninformative prior for $\alpha^k$ that may be used
generally when fitting the DDP-GP model. 

This approach works in practice because the parameter $\bbeta_0^k$
specifies the prior mean for the mean function of the GP prior, which
in turn formalizes the regression of $T^k$ on the covariates $\bx^k$,
including treatment selection. The imputed treatment effects hinge on
the predictive distribution under that regression. Excessive prior
shrinkage could smooth away the treatment effect that is the main
focus.  The use of an empirical Bayes type prior in the present
setting is similar to empirical Bayes priors in hierarchical models.
This type of empirical Bayes approach for hyperparameter selection is
commonly used when a full prior elicitation is either not possible or
is impractical.  Inference is not sensitive to values of the
hyperparameters $\blambda$ that determine the priors of $\sigma^k$ and
$\alpha^k$ for two reasons.  First, the
standard deviation $\sigma^k$ is the scale of
the kernel that is used to smooth the discrete random probability
measure generated by the DDP prior. It is important for 
 reporting a smooth fit, that is for display,
but it is not critical for the imputed fits in our
regression setting. Assuming some regularity of the posterior mean
function, smoothing adds only minor corrections.  Second, the total
mass parameter $\alpha^k$ determines the number of unique clusters
formed in the underlying Polya urn. Because most clusters are small
{\it a priori}, including many singleton clusters, varying the number
of these clusters by changing the prior of $\alpha^k$ does not
significantly change the posterior predictive values that are the
basis for the proposed inference. 

The conjugacy of the implied multivariate normal
on $\{\theta_h^k(\bx^k), h=0, 1, \dots\}$ and the normal kernel in
\eqref{eq:Fx} greatly simplify the computations, since any Markov chain Monte Carlo
(MCMC) scheme for DP mixture models can be used.
\cite{maceachern1998estimating} and
\cite{neal2000markov} described specific algorithms to implement
posterior MCMC simulation in DPM models. 
\cite{ishwaran2001gibbs} developed alternative computational algorithms
based on finite DPs, which 
truncated \eqref{eq:dpm} after a finite number of terms. 
We provide details of MCMC computations in the online supplement.

\section{Simulation Studies}
\label{sec:simu}
We conducted three simulation studies to evaluate the performance of
the proposed DDP-GP model as a tool for estimating the mean of $T$ in
survival regression settings.  The studies focused, respectively, on
estimation of
(i) survival regression;
(ii) regime effects in a study with two treatment arms and
single-stage regimes; 
and 
(iii) regime effects in a study with eight multi-stage regimes.  
For each of the latter two studies, the treatment assignment
probabilities depend on patient covariates.  
That is, we introduce a treatment selection bias. 
In all three simulations, we implement inference under DDP-GP models.
In (i) we use a single survival regression $F(Y_i \mid \bx_i)$ for a
patient-specific baseline covariate vector $\bx_i$.
For (ii) we still use a single DDP-GP model $F(Y_i \mid \bx_i, Z_i)$,
now adding a treatment indicator $Z_i$  to the survival regression. In
(iii) we use independent DDP-GP models $F^k(Y_i^k \mid \bx_i^k)$ for
multiple transition times, $k=1,\ldots,\ntrans$, similar to the
motivating application. 
For all three simulation studies, the 
hyperprior parameters were determined using the empirical Bayes approach
described earlier.  For all posterior computations, the MCMC algorithm
was implemented with an initial burn-in of 2,000 iterations and a total of
5,000 iterations, thinning out in batches of 10.  
This worked well in all cases,
with convergence diagnostics using the R package {\em coda} 
showing no evidence of  practical convergence problems.
Traceplots and empirical autocorrelation plots (not shown) for the
imputed parameters indicated a well mixing Markov chain.

\subsection{Survival Time Regression}
The first simulation was designed to study the DDP-GP regression model
by comparing inference for a survival function with the simulation
truth. In this study, we did not evaluate  a regime effect, but rather focused  on inference for the survival curve.

For each subject, we generated $T$ = survival time,  the
covariates $x_1$ = tumor size (0=small,
1=large)  and $x_2$ = body weight, and $x_3$ =  a biomarker (0=absent, 1=present).
We assumed that small and large tumor sizes each had probability .50.
Body weights were computed by sampling from a uniform distribution, $\mathrm{Unif}(80, 150)$, with the covariate $x_2$ defined by shifting and scaling to obtain mean 0 and variance 1.
The biomarker was associated with  tumor size, as follows.  Patients in the large tumor size group
were biomarker negative  with probability 0.7 and biomarker positive with probability
0.3. Patients with small tumor size were  biomarker negative with probability 0.3 and  biomarker positive
with probability 0.7.
Let $Y \sim \LN(m,s)$ denote a log normal random variable
$Y=\log T$ for $T \sim \N(m,s)$. By a slight abuse of notation, we
also use $\LN(m,s)$ to denote the log normal p.d.f.
 Let $\bx_i=(1, \mathrm{x}_{i,1}, \mathrm{x}_{i,2},\mathrm{x}_{i,3})$
denote the covariates for patient $i$.
We simulated each sample $Y_1,\cdots,Y_n$ of $n$ observations from a mixture
of lognormal distributions, $Y_i|\bx_i\ \sim\  0.4 \ \LN(\bx_i\bbeta_1,
\sigma^2)+0.6 \ \LN(\bx_i\bbeta_2, \sigma^2)$,
where the true covariate parameters of the mixture components were
 $\bbeta_1=(1, 2, -2, 1)'$ and $\bbeta_2=(2, -1, 3,
-3)'$, with $\sigma^2=0.4$.
For comparison, we also fit an 
AFT regression model, assuming
 $$Y_i\ =\ \log(T_i)=\bx_i'\bbeta+\sigma\epsilon_i, \ \ \ i=1,\dots, n
$$
with $\epsilon_i$ following an extreme value distribution,
so that   $T_i$ follows a Weibull distribution.

\begin{figure}[!htbp]
\centering
\begin{tabular}{cc}
\includegraphics[scale=0.45]{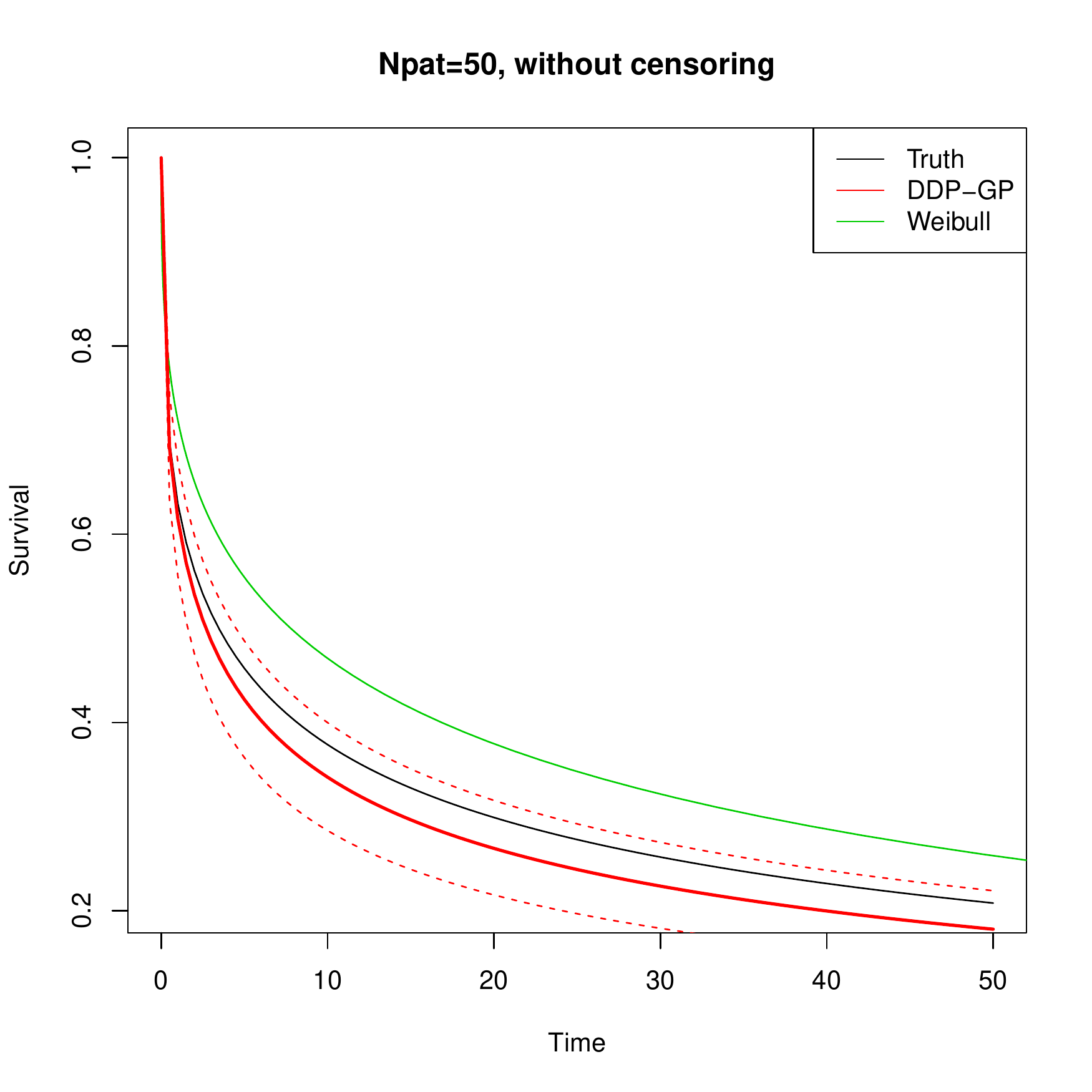}&\includegraphics[scale=0.45]{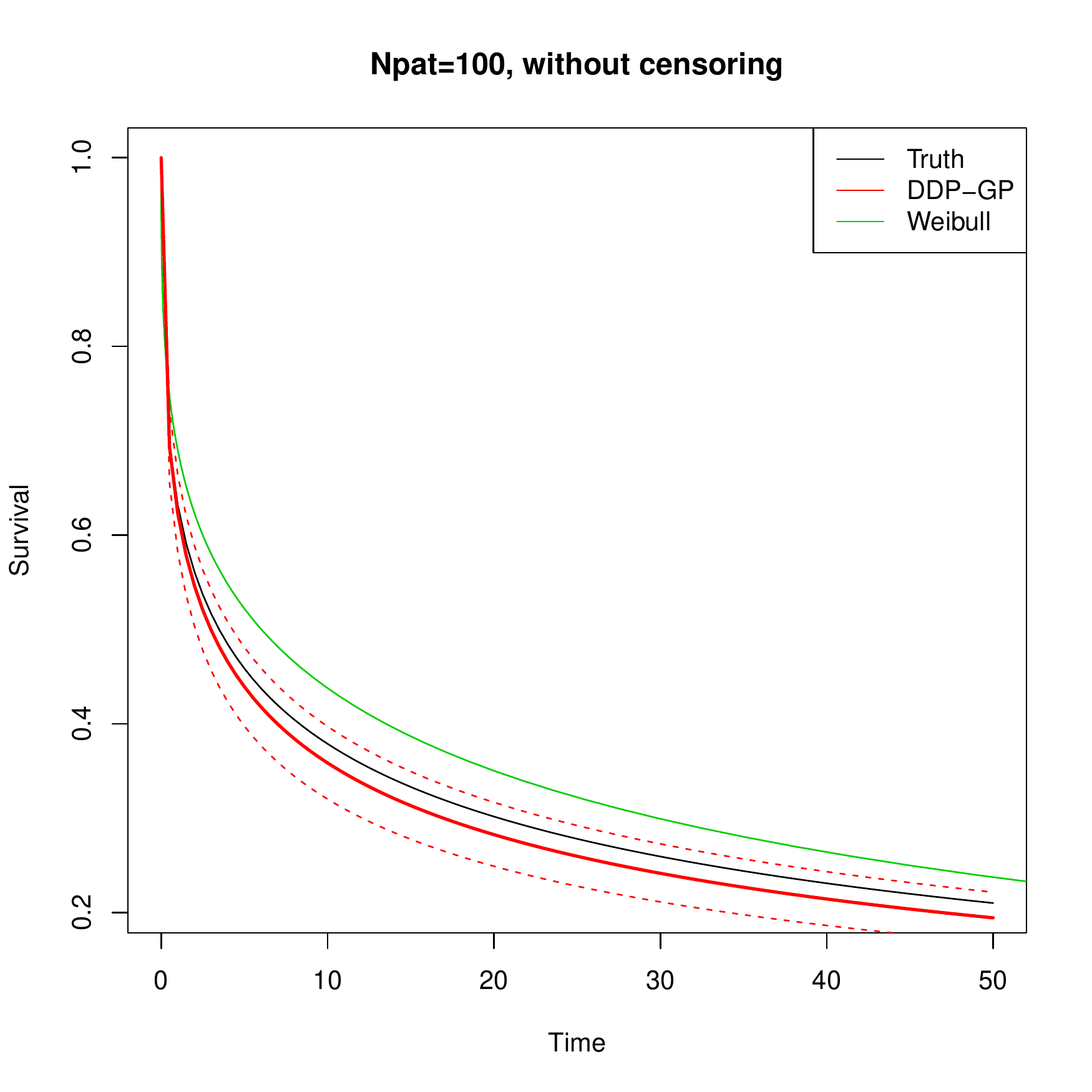}\\
\includegraphics[scale=0.45]{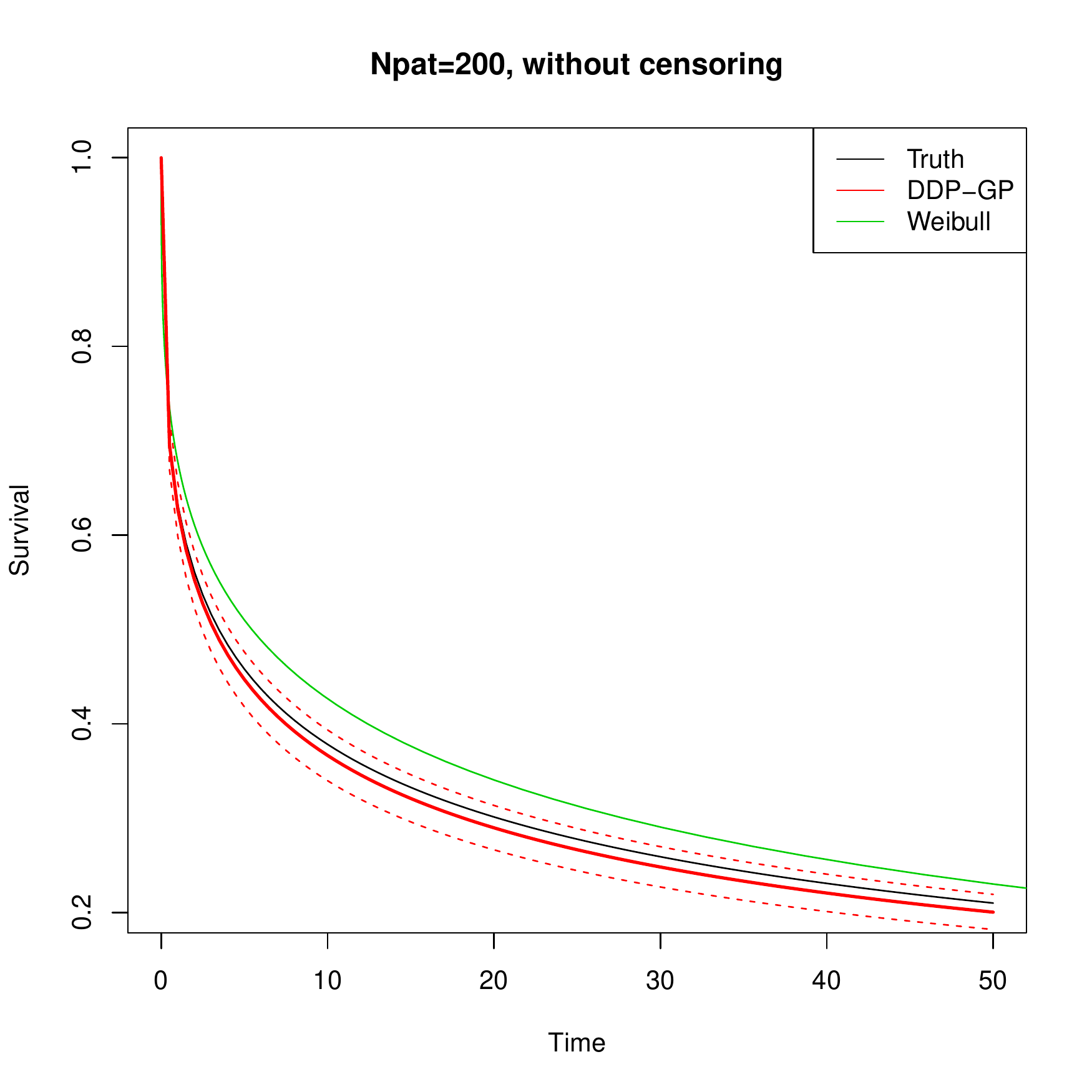}&\includegraphics[scale=0.45]{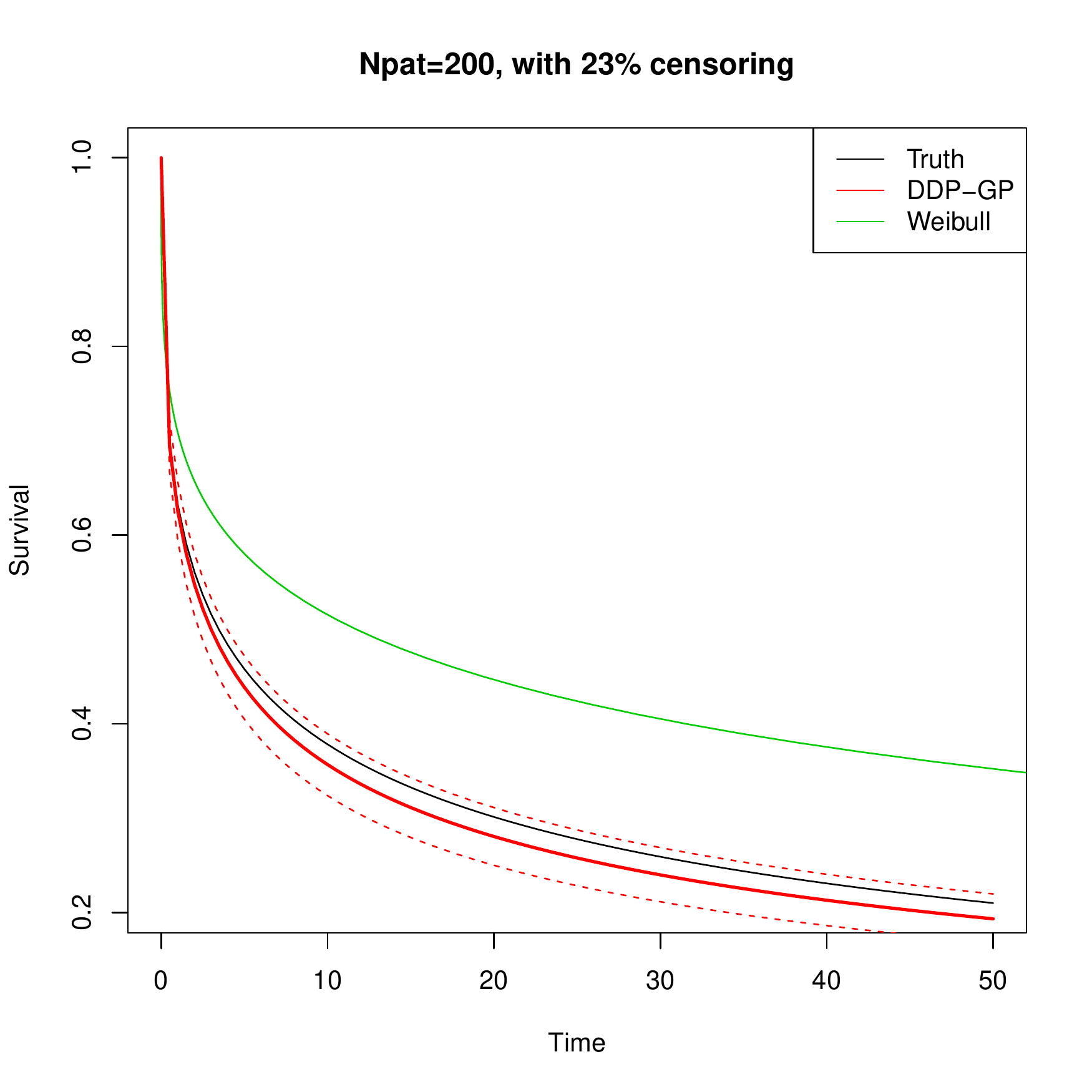}\\
\end{tabular}
\caption{{\small Simulation example 1. True mean survival functions (black color) and estimated mean survival functions under the DDP-GP model (red color) for sample sizes $n=50, 100, 200$ and $n=200$ with 23\% censoring for 1,000 simulations.  For comparisons, we also show the MLE under an AFT regression with Weibull distribution (green color). In all cases, the point-wise 90\% credible bands are also displayed as the region between two dotted red lines. }}
\label{fig:simu1}
\end{figure}

In this simulation, we considered four scenarios, with $n=50, 100,$ or $200$
observations without censoring or $n=200$ with 23\% censoring.
For each scenario,  $N=1,000$ trials were simulated.
For each simulated data set we fit a DDP-GP survival regression model
$F(Y_i \mid \bx_i)$.
For simulation $j$, let $\Sbar(t \mid \bx) = 
p(T_{n+1} \geq t \mid \bx_{n+1,j}=\bx, data)$ denote
the posterior expected survival function for a future patient with
covariate $\bx$.
Using the empirical distribution 
$\frac1n \sum_{i=1}^n \delta_{\bx_{ij}}$ to marginalize
w.r.t. $\bx_{n+1,j}$ and averaging across simulations, we get
$$
  \Sbar(t)  =
  \frac{1}{N}\,\sum_{j=1}^N\,\frac1n \sum_{i=1}^n\Sbar(t\mid \bx_{ij}).
$$
Figure \ref{fig:simu1} compares $\Sbar(\cdot)$ under 
the DDP-GP model with the simulation truth 
$$
S_0(t) = \frac{1}{N}\,\sum_{j=1}^N\,\frac1n\, \sum_{i=1}^nS_0(t\mid \bx_{ij}),
$$
and a maximum likelihood estimate (MLE) under a Weibull AFT model. 
In each scenario, the true curve is given  as a solid black solid line,
the MLE of the survival functions under the AFT regression model assuming a  Weibull distribution  as  a solid green solid line,
and the posterior mean survival function under the DDP-GP model  as a solid red
line with point-wise 90\% credible bands as two dotted red lines.

In all four scenarios, the DDP-GP model based estimate reliably recovered the shape of
the true survival function and avoided the excessive bias seen with the Weibull
MLE.
As expected,  the three scenarios without
censoring show that increasing  sample size gives more accurate estimation.
With 23\% censoring, the DDP-GP estimate becomes less accurate,
but it still is much closer to the simulation truth than the Weibull MLE.

\subsection{Estimating a Treatment Effect in Single-stage Regimes}
The second simulation study was designed to investigate 
inference under the DDP-GP model for a regime effect
in a single-stage treatment setting.
The simulated data represent  what might be obtained in an
  observational setting where treatment is chosen by the attending physician based
on patient covariates, rather than from a fairly randomized clinical trial.
We simulated a binary treatment indicator $Z_i \in$
\{0=control,\ 1=experimental\} that depended on
two continuous covariates, 
$\bx_i=(L_i, W_i)$, for $n=100$ patients, $i=1, \dots, n$.
For example, $L_i$ could be a patient's creatinine to quantify kidney function, and
$W_i$ could be body weight.
We generated $L_i$ from a mixture of normals,
$L_i \sim \frac12 N(40,10^2) + \frac12 N(20,10^2)$,  which could correspond to a subgroup of patients having worse kidney function (higher creatinine level) due to damage from prior chemotherapy.
We assumed that $W_i \sim \Unif(-\sqrt{12},\sqrt{12})$, a uniform
with zero mean and unit standard deviation, as could
arise from standardizing a uniformly distributed raw variable.
We generated the treatment indicators using the modified logistic regression model
{\small
\[ p(Z_i = 1 \mid L_i, W_i)= \left\{
  \begin{array}{l l}
    0.05 & \quad \text{if $\left\{1+\exp[-2(L_i-30)/10] \right\}^{-1}\leq 0.05$ }\\
    0.95 & \quad \text{if $\left\{1+\exp[-2(L_i-30)/10] \right\}^{-1}\geq 0.95$}\\
     \left\{1+\exp[-2(L_i-30)/10] \right\}^{-1} & \quad \text{otherwise},
  \end{array} \right.\]}
that is, a logistic regression with intercept 30 and slope 1/5  truncated at 0.05 and 0.95.
This produces  a very unbalanced treatment assignment, for example,
$p(Z_i=1 \mid L_i=40) = 0.88$ versus $p(Z_i=1 \mid L_i=20)=0.12$.
This could arise  in a setting where standard
therapy, $Z=0$, is known to be nephrotoxic, while it is believed by most of the treating physicians that
the experimental therapy, $Z_i=1$, is not, so patients with high creatinine are more likely to
be given the experimental therapy.
In this simulation study, the goal is to estimate the comparative effect on survival of the
experimental therapy versus the control.
In the two treatment arms, we generated patients' responses from 
$$Y(1)\sim \frac{1}{2} \ \mathrm{N}(3-0.2L+\sqrt{L}-0.1W,\
\sigma)+\frac{1}{2} \ \mathrm{N}(2-0.2L+\sqrt{L}-0.1W,\ \sigma)$$ and $$Y(0)\sim
\mathrm{N}(-0.2L+\sqrt{L}-0.1W,\ \sigma),
$$ 
with $\sigma = 0.4$. We simulated 1,000 trials.
Note that under the simulation truth the treatment effect, $E[Y(1)-Y(0) \mid \bx=(L,W)]=2.5$, is constant across $L, W$.

\begin{figure}[!h]
\centering
\begin{tabular}{cc}
\includegraphics[scale=0.45]{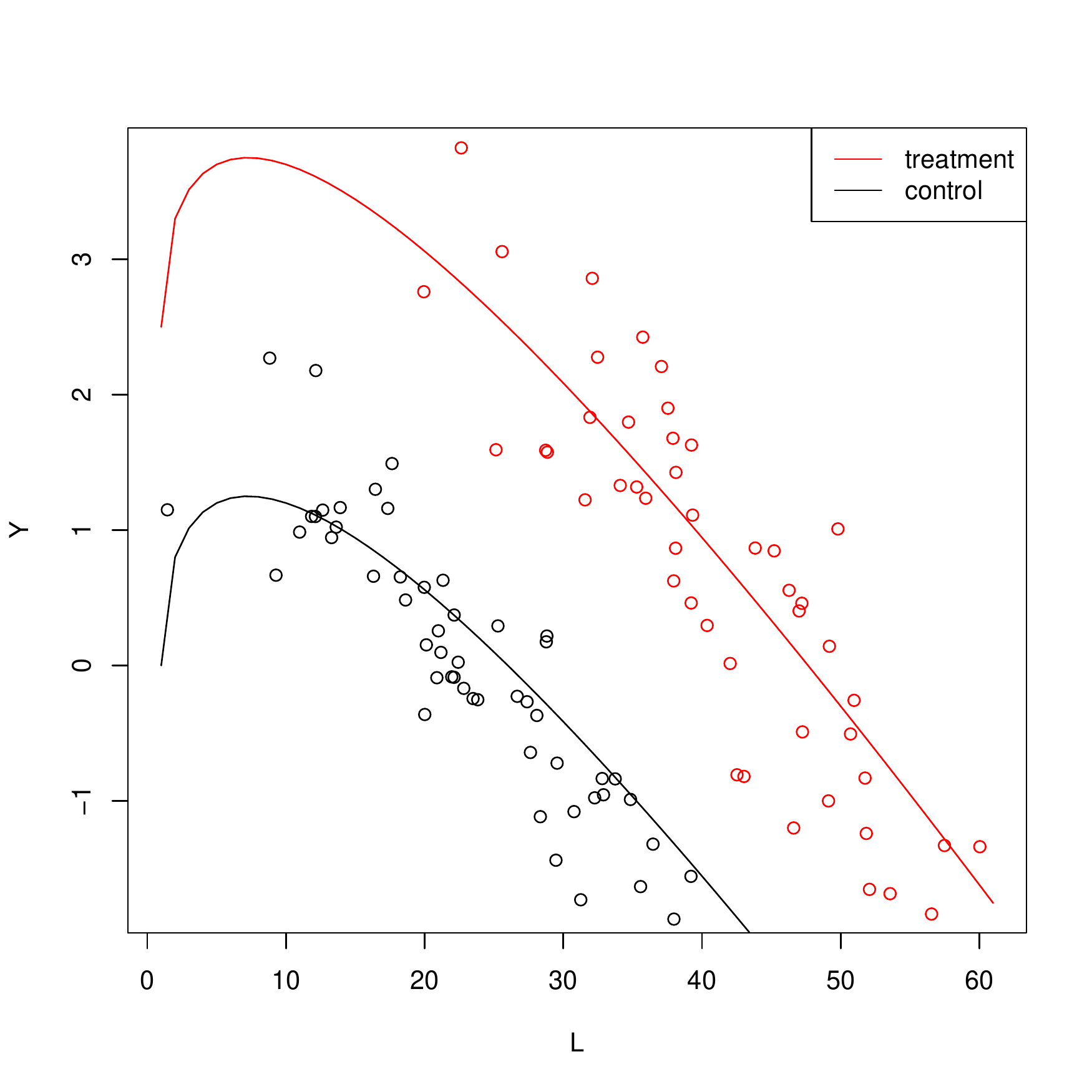}&\includegraphics[scale=0.45]{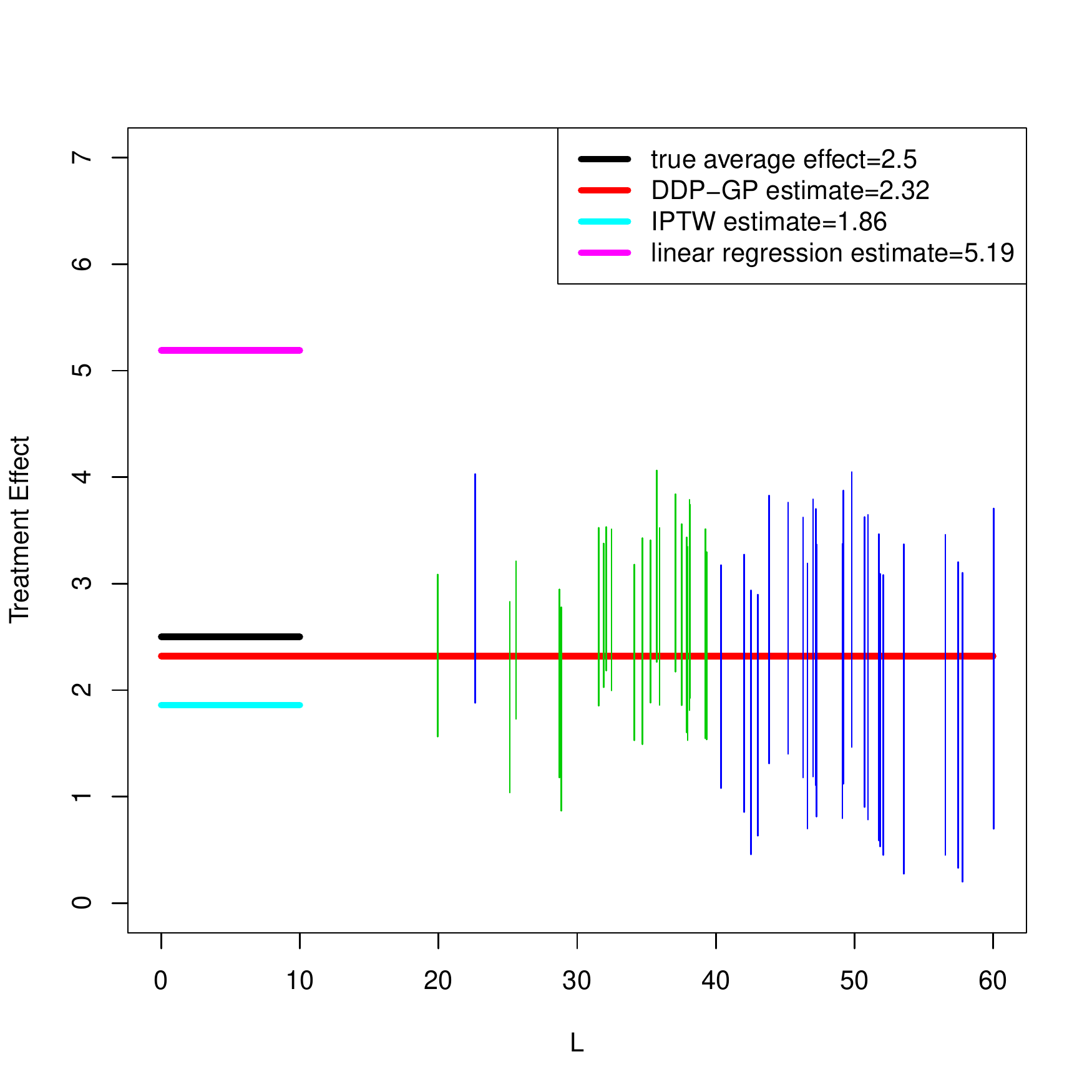}\\
(a) & (b)\\
\end{tabular}
\caption{{\small Simulation example 2.
  (a) Simulated data for one (treatment, control) pair. The upper red solid curve represents $E[Y(1)\mid
  X]$, the lower black curve represents $E[Y(0)\mid X]$ given $W=0$. The red dots
  close to the upper curve are the treated observations and the black
  dots close to the lower curve are the untreated.
  (b) Average treatment effect estimations. The black solid line represents
  the true average treatment effect, the red line represents posterior mean
   treatment effect estimates under the DDP-GP model, turquoise blue
  represents IPTW estimate, the heliotrope line represents the linear regression
  estimate. The  vertical line segments are marginal 90\%
  posterior intervals for the treatment effect at each $L$ value from
  treated observations.  }}
\label{fig:simu2}
\end{figure}

Figure \ref{fig:simu2}(a) plots the simulation truth for the mean
response curve under $Z=1$ and $Z=0$ versus $L$, with $W \equiv 0$, in
one randomly selected trial.  The upper red solid curve represents
$E[Y(1)\mid L, W=0]$ and the lower black curve represents $E[Y(0)\mid
L, W=0]$. The red dots close to the upper curve are the observations
for experimental arm patients and the black dots close to the lower
curve are the observations for the control arm patients.  
We define an average treatment effect 
 for the entire population under the simulation truth
as $\frac{1}{n}\sum_{i=1}^nE[Y_i(1)-Y_i(0)]=2.5$.

We implemented inference for a survival regression 
$F(Y_i \mid \bx_i, Z_i)$ using the proposed DDP-GP model.
Figure \ref{fig:simu2}(b) summarizes inference for the data from panel
(a). 
Let $\hat Y_i(z) = E(Y_{n+1} \mid L_{n+1}=L_i, W_{n+1}=W_i, Z_i=z,
data)$ denote the posterior expected response for a future patient
$n+1$.
We define an estimated average treatment effect as 
$\frac1n \sum_{i=1}^n [\hat Y_i(1) - \hat Y_i(0)]$.
Figure \ref{fig:simu2}(b) shows the estimated average treatment
effect (horizontal red line), and credible intervals 
for individual effects $\hat Y_i(1)-\hat Y_i(0)$ (vertical line
segments, located at $L_i$).

For comparison, we also applied both linear regression (LR) and an IPTW method
to the simulated data to estimate the average treatment effect. The
LR method fits observations from both treatments
using linear predictor functions and estimates the average treatment
effect,  assuming
$Y_i(1)=\beta_{10}+\beta_{11}L_i+\beta_{12}W_i+\epsilon_{1i}$ and
$Y_i(0)=\beta_{00}+\beta_{01}L_i+\beta_{02}W_i+\epsilon_{0i}$.
Denoting the least squares estimates by  $\hat{\beta}_{zj}$
for $z=0,1$ and $j=0,1,2,$ the estimated means are
$\hat{E}\{Y_i(z)\} =
\hat{\beta}_{z0}+\hat{\beta}_{z1}L_i+\hat{\beta}_{z2}W_i$.
We define an estimated average treatment effect as
$
   \frac1n \sum_i\,[\hat{E}\{Y_i(1)\}-\hat{E}\{Y_i(0)\}]
$.
The IPTW method assigns each patient $i$ a weight $b_i$ equal to the
inverse of  an estimate of $p(Z_i\mid \bx_i)$, the conditional probability of receiving his or her actual
treatment \citep{robins2000marginal}, with the estimate obtained by fitting
 a logistic regression model.  The effect of weighting is to
create a pseudo-population consisting of $b_i$ copies of each patient
$i$. For example, if $b_i=5$ then  five copies of the $i^{th}$ patient
are contributed to the pseudo-population. Thus, for $z$ = 0 or 1,
we define an estimated mean outcome 
$$
\IPTW(Z=z)=\frac{\sum_iI(Z_i=z)b_iY_i}{\sum_iI(Z_i=z)b_i},
$$
and a corresponding average treatment effect estimate
 $
 \IPTW(Z=1)-\IPTW(Z=0).
 $
The DDP-GP point estimate  of the average effect of
the treatment is the posterior mean 2.31 with 90\% posterior credible interval (1.89, 2.96). The
LR fit yields an overestimate, 4.13, while IPTW yields
an underestimate, 1.11.  In Figure \ref{fig:simu2}(b), the red
horizontal line represents the posterior mean  treatment effect
estimate obtained from the DDP-GP model. The short
horizontal black, turquoise blue and heliotrope
solid lines represent the true average treatment effect, IPTW estimate, and
LR estimate, respectively.  The vertical green and blue
segments are marginal 90\% posterior credible intervals for the treatment
effect at each $L$ value from treated observations. Lengths of
posterior  credible intervals larger than 2 are highlighted by blue
segments. Note that the uncertainty bounds grow wider in the range where there is 
 less overlap across 
treatment groups,  that is, over a range of covariate values 
for which we do not observe reliable  
empirical counterfactuals for each data point (e.g. $L>50$). Most
of the credible intervals reasonably cover the true treatment effect.

\begin{figure}[!]
\centering
\includegraphics[width=10cm, height=8cm]{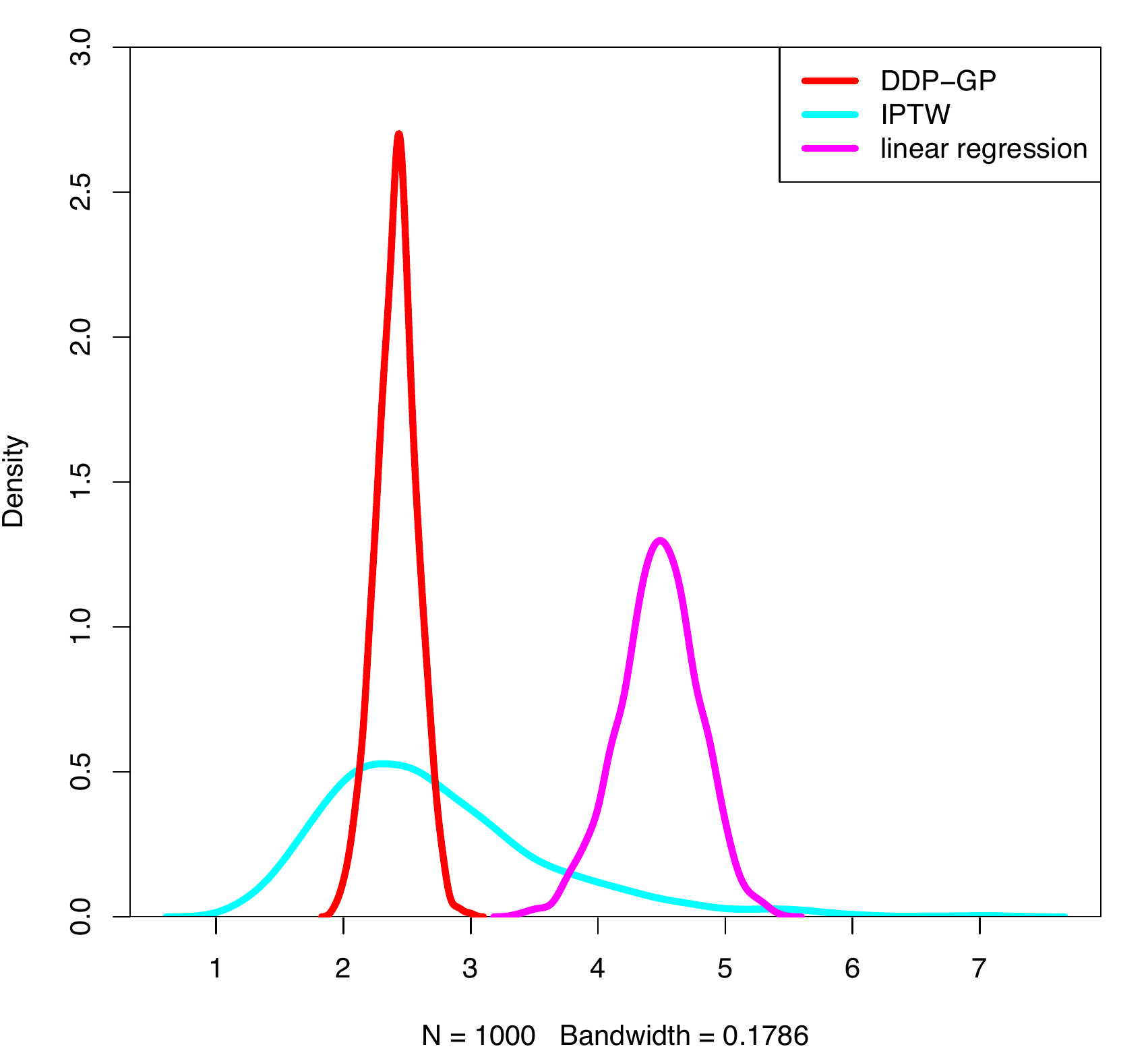}
\caption{The density plot of estimated regime effects by DDP-GP, IPTW and linear regression in 1,000 trials.}
\label{fig:density}
\end{figure}

Figure \ref{fig:simu2} reports inference for one hypothetical
data set. For a more meaningful comparison we carried out extensive
simulation and report the distribution of estimated 
regime effects across repeat simulations. 
We compared the regime effects estimates obtained by DDP-GP, IPTW, and
LR based on data from 1,000 simulated trials. Figure \ref{fig:density}
gives density plots of the estimated regime effects.  Compared to the
estimates obtained from DDP-GP, the IPTW estimates are much more
variable, ranging from 1.14 to 7.13.  In general, the LR estimates are
highly biased, and overestimate the true effects.
The distribution of estimated regime effects under the DDP-GP model is
remarkably narrowly centered around the simulation truth, in
comparison with the two alternative methods. 

\subsection{Regime Effect for Multi-stage Regimes}
Our third simulation study was designed to examine inference on
strategy effects for multi-stage regimes.  This simulation is a
stylized version of the leukemia data that we will analyze in
Section 6.  We simulated samples of size $n=200$.  Patients initially
were randomized between two induction therapies, with the
randomization probabilities based on their blood glucose values, which
were simulated as $L_i\ \sim\ \N(100, 10^2)$.  Denoting $Z^1\in \{a_1,
a_2\}$, if $L_i<100$, then $Z_i^1=a_1$ with probability 0.6 and
$Z_i^1=a_2$ with probability 0.4. If $L_i\geq 100$, then $Z_i^1=a_1$
with probability 0.4 and $Z_i^1=a_2$ with probability 0.6.
We then generated a response (see below). 
For patients who were resistant ($R$) to their induction therapies,
they were assigned salvage treatment $Z^{2,1}\in\{b_{11},
b_{12}\}$. If their blood glucoses were smaller than 100,
$Z^{2,1}=b_{11}$ with probability 0.8 and $Z^{2,1}=b_{12}$ with
probability 0.2; if their blood glucoses were larger than 100,
$Z^{2,1}=b_{11}$ with probability 0.2 and $Z^{2,1}=b_{12}$ with
probability 0.8. Patients who achieved $C$ and subsequently suffered
disease progression ($P$), were given salvage treatment $Z^{2,2}\in
\{b_{21}, b_{22}\}$. The salvage treatment for each patient
$Z_i^{2,2}$ was assigned based on his/her baseline covariate $L_i$: if
$L_i<100$, $Z_i^{2,2}=b_{21}$ with probability 0.2 and
$Z_i^{2,2}=b_{22}$ with probability 0.8; if $L_i\geq 100$,
$Z_i^{2,2}=b_{21}$ with probability 0.85 and $Z_i^{2,2}=b_{22}$ with
probability 0.15.  Thus, the survival time for each patient was
evaluated as 
$$
  T_i = \left\{ \begin{array}{ll}
   T_i^{(0, R)}+T_i^{(R, D)} &\mbox{ if patient $i$ had sequence $(L, Z^1, T^{ (0, R)}, Z^{2, 1})$}\\
   T_i^{(0, C)}+T_i^{(C, P)}+T_i^{(P, D)} &\mbox{ if patient $i$ had sequence $(L, Z^1, T^{(0, C)}, T^{(C, P)}, Z^{2, 2})$}.
         \end{array} \right.
$$
We simulated the times of two completing risks $R$ and $C$ as  $T_i^{(0, R)} \sim LN(\bbeta^{(0, R)}\bx_i^{(0, R)}, \sigma^{(0, R)})$ and $T_i^{(0, C)} \sim LN(\bbeta^{(0, C)}\bx_i^{(0, C)}, \sigma^{(0, C)})$, where $\bbeta^{(0, R)}=(2,\, 0.02,\, 0)$, $\bbeta^{(0, C)}=(1.5,\, 0.03,\, -0.8)$, with $\bx_i^{k}=(1, L_i, Z_i^1)$ for $k\in \{(0, R), (0, C)\}$.
For  transitions
$k\in\{ (R, D), (C, P), (P, D)\}$, we generated transition times $T_i^k \sim LN(\bbeta^k\bx_i^k, \sigma^k)$, where $\bbeta^{(R, D)}=(-0.5,\, 0.03,\, 0.2, 0.5,\, 0.3)$, $\bbeta^{(C, P)}=(1,\, 0.05,\, 1,\, -0.6)$, 
$\bbeta^{(P, D)}=(0.8,\, 0.04,\, 1.5,\, -1,\, 0.5,\, 0.5)$, with covariate vectors \\  $\bx_i^{(R, D)}=(1,\, L_i,\, Z_i^1, \log(T_i^{(0, R)}), Z_i^{2,1})$,
$\bx_i^{(C, P)}=(1, L_i, Z_i^1, \log(T_i^{(0, C)}))$ and
$\bx_i^{(P, D)}=(1, L_i, Z^1_i,$ $\log(T_i^{(0, C)}),$ $\log(T_i^{(C, P)}), Z^{2,2}_i)$. We simulated $N$ = 1,000 trials with 15\% censoring.

\begin{figure}[!h]
\centering
\includegraphics[width=.9\textwidth]{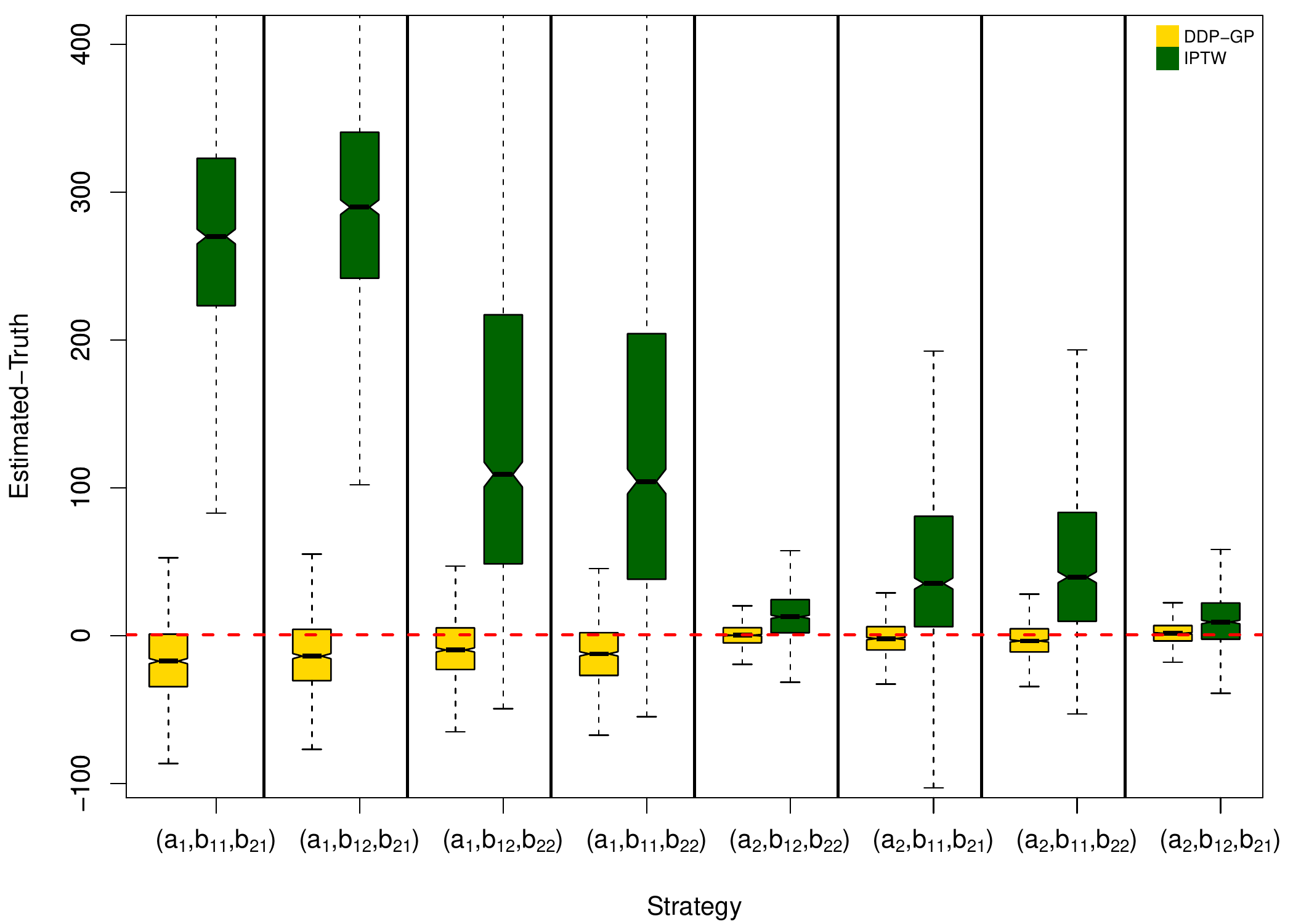}
\caption{Simulation 3. Small (yellow) boxplots show posterior estimated OS under each of the 8 regimes as a difference with the simulation truth over 1,000 simulations. The large (green) boxes show inference under the IPTW approach. In each notched box-whisker plot, the box shows the interquartile range (IQR) from 1st quantile ($Q1$) to 3rd quantile ($Q3$), and the mid-line is the median. The top whisker denotes $Q3$+$1.5*$IQR and the bottom whisker  $Q1$-$1.5*$IQR.  The notch displays a confidence interval for the median, that is median$\pm 1.57*IQR/\sqrt{n}$. Here $n=1000$. }
\label{fig:sim-DTR}
\end{figure}

The goal is to estimate mean survival time for each DTR $(Z^1, Z^{2,1}, Z^{2,2})$. We have 8 possible DTRs in this simulation. We applied both inference under the Bayesian nonparametric DDP-GP model and IPTW to
the each simulated dataset to estimate mean survival for each of the
eight possible DTRs. 
For the nonparametric Bayesian inference we defined independent
DDP-GP models $F^k(Y_i^k \mid \bx_i^k)$
for each of the $\ntrans=5$ log transition times $Y_i^k=\log T_i^k$.
Figure \ref{fig:sim-DTR} gives comparisons of the
mean survival estimates using boxplots  of (Estimated mean survival -
Simulation truth), based on the simulation sample of 1000 datasets, 
obtained by  DDP-GP and IPTW,  for each possible DTR. The yellow
boxplots represent the DDP-GP posterior mean estimates and the green
boxplots represent the IPTW estimators. Figure \ref{fig:sim-DTR} shows
that the DDP-GP  estimates  on average are much closer to  the truth
and have much smaller variability,  compared to the IPTW estimates, across all eight scenarios.

\section{Evaluation of the Leukemia Trial Regimes}

\subsection{Computing Mean Survival Time}

We first review the likelihood used by \cite{wahed2013evaluating} as a basis for
frequentist G-estimation of mean survival time for the leukemia trial
regimes.  We will apply the Bayesian nonparametric DDP-GP model to
this basic structure to obtain posterior means and credible intervals
of mean survival time for each DTR.

Recall that the disease states are $D$ = death, $R$ = resistant disease,
$C$ = complete remission, and $P$ = progressive disease. In stage $\ell=1$
(induction chemotherapy), the three events $D,$ $R,$ and $C$ are
competing risks since only one can be observed. For the $i^{th}$
patient, the stage 1 outcome is $s_{1i}=D$ if the patient dies,
$s_{1i}=R$ if the patient's disease is resistant to induction, and $s_{1i}=C$ if
induction achieves CR. The corresponding transition times  are
$T_i^{(0,D)}$  =  time to D
(the left most arrow in Figure \ref{fig:flowchart}),
 $T_i^{(0,R)}$ = time to R, and $T_i^{(0,C)}$ = time to C.
In stage 2,  the  transition time $T_i^{(R,D)}$ is defined only if $s_{1i}=R$,
$T_i^{(C,D)}$ is defined only if $s_{1i}=C$ and
$s_{2i}=D$,  and  $T_i^{(C,P)}$ is defined only if $s_{1i}=C$  and $s_{2i}=P$.
 The time from post-CR progression to death,
 $T_i^{(P,D)}$, is defined if $s_{1i}=C$ and $s_{2i}=P$.
We thus define seven counterfactual transition times $T_i^k$,
where $k$ indexes the transitions  $(0,D), (0,R), (0,C), (R,D),
(C,D), (C,P), (P,D)$. 
Figure \ref{fig:flowchart} shows a flowchart of the possible  outcome pathways.
A dynamic treatment regime for this data may be expressed as  ${\bf Z}$ = $(Z^1, Z^{2,1}, Z^{2,2})$
where $Z^1$ is the induction chemo, $Z^{2,1}$ is the salvage therapy given if $s_{1i}=R$,
and  $Z^{2,2}$ is the salvage therapy given if $s_{1i}=C$ and $s_{2i}=P.$

Our primary goal is to estimate mean survival time for each DTR
$(Z^1, Z^{2,1}, Z^{2,2})$  while accounting for baseline covariates and
non-random treatment assignment.  Under the DDP-GP model,
we denote the mean survival time for
a future patient under $\bZ$ by  
\begin{equation}
  \eta(\bZ)=E(T\mid \bZ).
\label{eq:etaZ}
\end{equation}
The survival time for a future patient $i=n+1$ is 
\begin{multline}
T_i = I(s_{1i}=D)T_i^{(0,D)}+I(s_{1i}=R)(T_i^{(0,R)}+T_i^{(R,D)})\\
      +I(s_{1i}=C)\{ I(s_{2i}=D)(T_i^{(0,C)}+T_i^{(C,D)})+
      I(s_{2i}=P)(T_i^{(0,C)}+T_i^{(C,P)}+T_i^{(P,D)})\}.
\label{eq:OS}
\end{multline}
The expectation of \eqref{eq:OS} under the DDP-GP model is evaluated
by applying the law of total probability, 
using the same steps as in \cite{wahed2013evaluating}. We first condition on the four possible cases, $(s_{1i}=D)$,
$(s_{1i}=R)$, $(s_{1i}=C, s_{2i}=D)$ and $(s_{1i}=C, s_{2i}=P)$,
compute the conditional expectation in each case,
and then average across the cases. 
This computation requires evaluating seven expressions for the
conditional mean transition times 
$$
  \eta^k({\bf Z}, \bx^k) = E(T^k \mid \bZ, \bx^k)
$$
under
$F^k(\cdot \mid \bx^k)$, for each $k$.
For example,
$\eta^{(P,D)}(Z^1, Z^{2,2}, \bx^0, T^{(0,C)}, T^{(C,P)})$
is the conditional mean remaining survival time, from $P$ to $D$,
given that  $C$ was achieved in stage 1 with frontline therapy $Z^1$,
followed by $P$ and salvage therapy $Z^{2,2}$ in stage $2$.
The DDP-GP models for $F^k(\cdot \mid \bx^k)$, $k=1,\ldots,\ntrans=7$ 
define most of the marginalization for the expectation
in $\eta(\bZ)$, leaving only conditioning on the baseline
covariates $\bx_i^0$. As \cite{wahed2013evaluating}, we use the empirical distribution 
$\phat(\bx^0)$ over the observed patients to define an overall
mean survival time \eqref{eq:etaZ}.
The described evaluation of $\eta(\bZ)$ is an application of
Robins's $G$-formula \citep{robins1986new, robins2000marginal}.  The
complete expression is given as equation \eqref{eq:likebased} 
in the Appendix. 
In the upcoming discussion we will use $\eta(\bZ)$ to evaluate
and compare the proposed approach.

\subsection{Leukemia Data -- Inference for the 
   Survival Regression}  
To analyze the AML-MDS trial data under the proposed DDP-GP model, we
first implement posterior inference for six of the $\ntrans=7$ transition
times. The exception is $T^{(C,D)}$.
Due to the limited sample size -- only 9 patients died after $C$ without first suffering disease progression ($P$) -- 
we do not implement the DDP-GP model, and instead use an
intercept-only Weibull AFT model.   
Table \ref{table:table1} summarizes the data. The table reports
the number of patients and median transition times for 
some selected transitions. 

\begin{figure}[!htbp]
\centering
\begin{tabular}{cc}
\includegraphics[width=8cm,height=8cm]{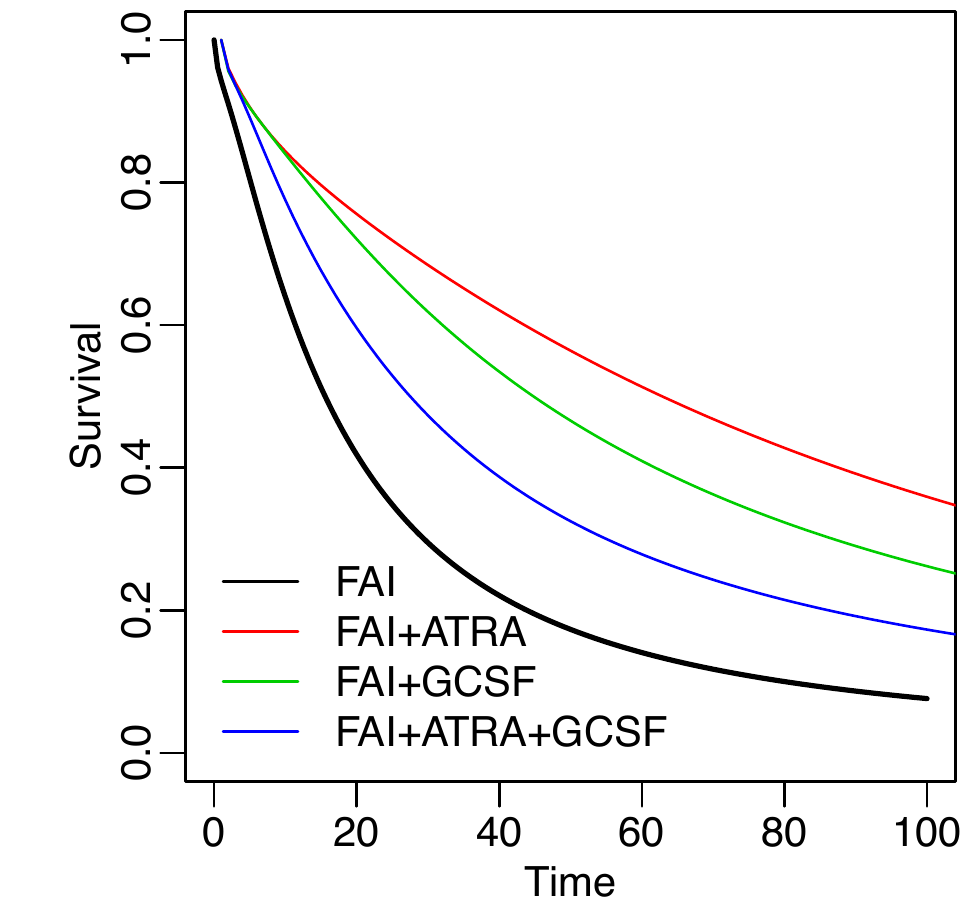}&
\includegraphics[width=8cm,height=8cm]{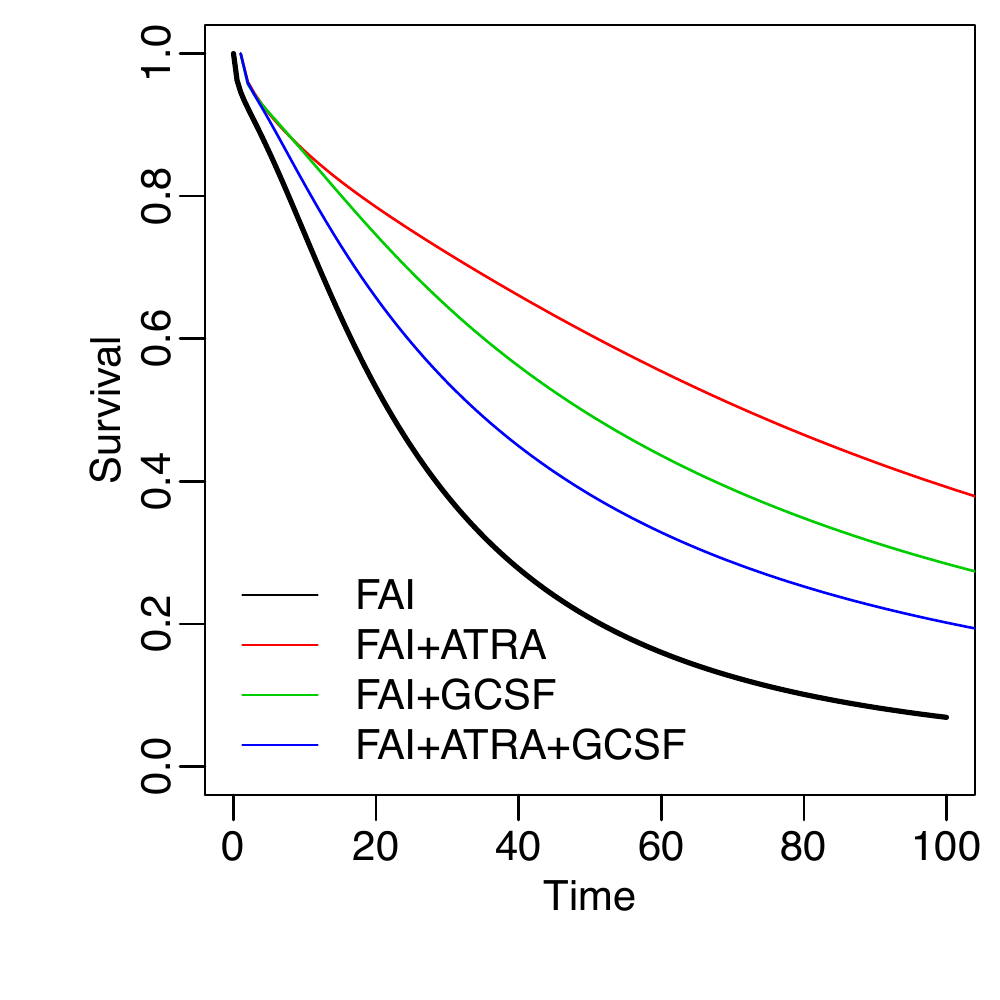}\\
(a) $Z^{2,1}=$ HDAC, $T^{(0,R)}=20$ &
(b) $Z^{2,1}=$ non-HDAC, $T^{(0,R)}=20$ \\
\includegraphics[width=.5\textwidth]{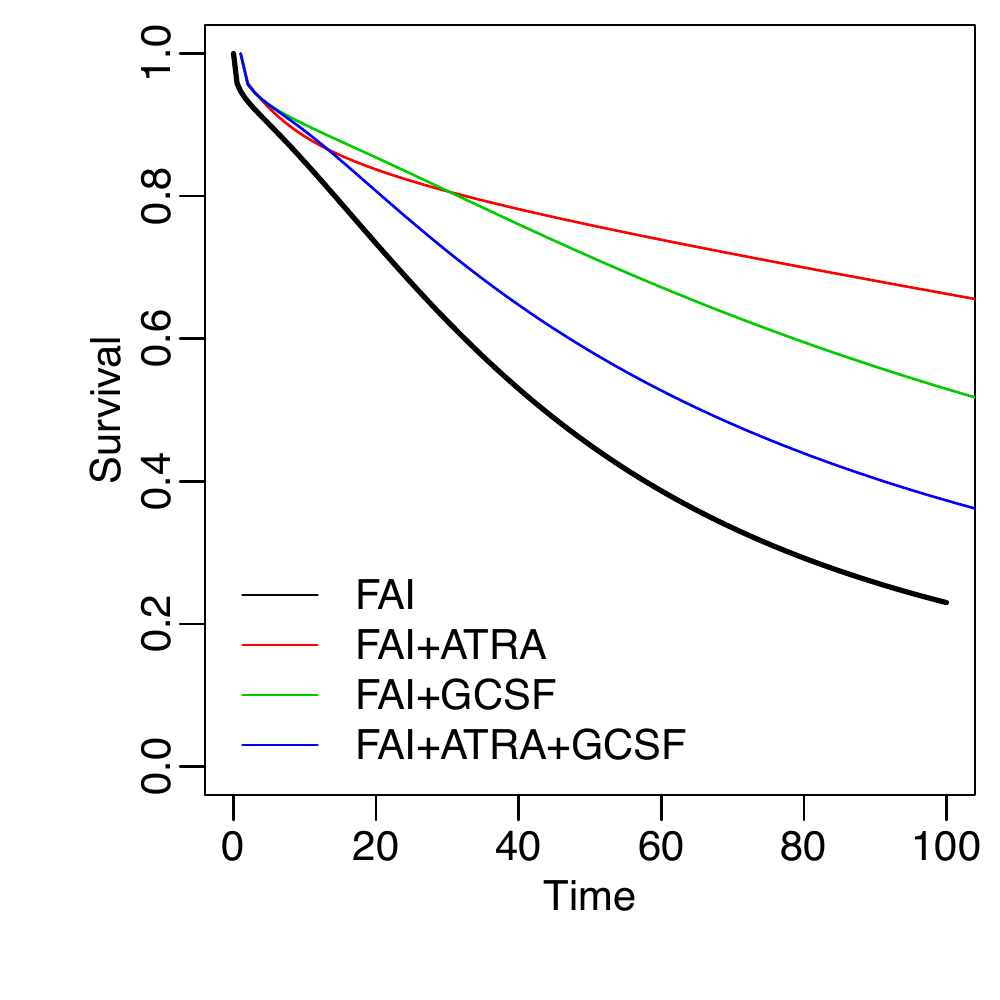}&
\includegraphics[width=.5\textwidth]{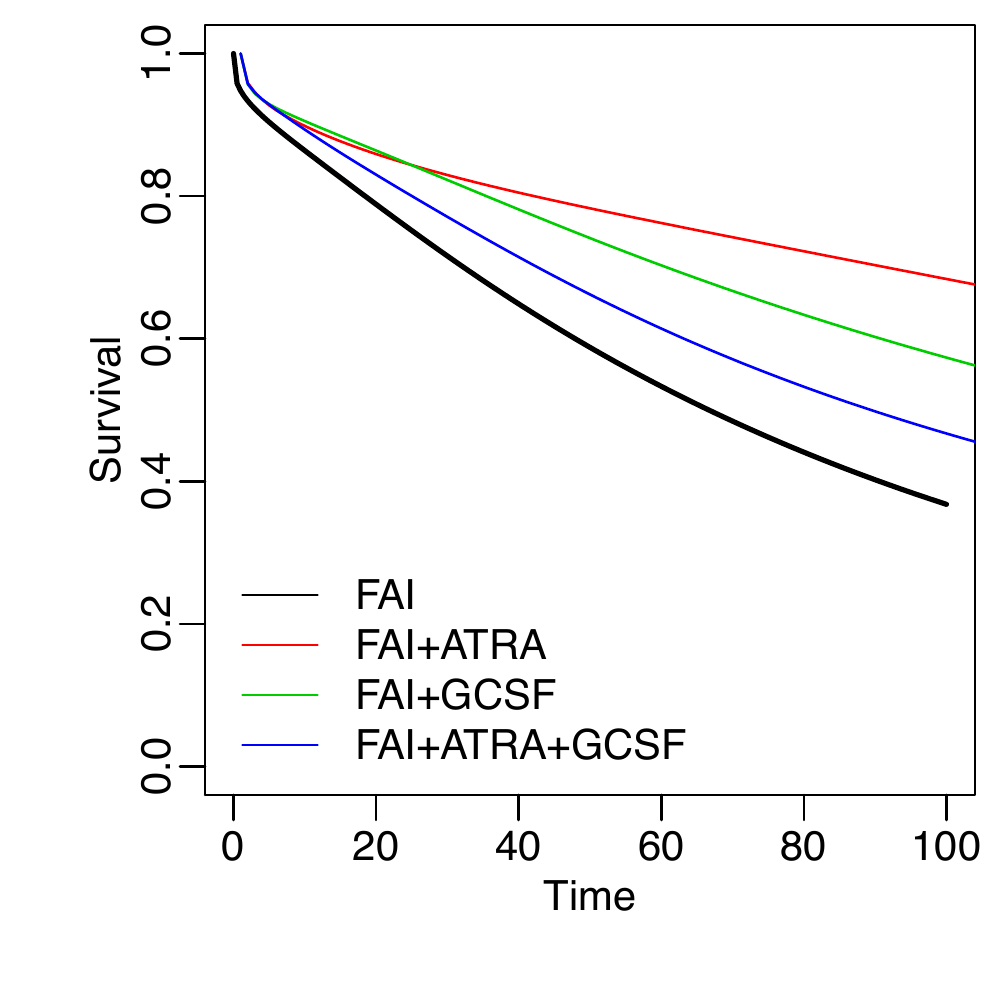}\\
(c) $Z^{2,1}=$ HDAC; $T^{(0,R)}=55$ &
(d) $Z^{2,1}=$ non-HDAC; $T^{(0,R)}=55$ 
\end{tabular}
\caption{{\small Survival regression for
    $T^{(R,D)}$ in the AML-MDS trial.
    Panels (a)-(d) show the posterior estimated survival functions 
    for a future patient at scaled age 0 with poor prognosis cytogenetic
    abnormality, with $T^{(0,R)}$ and $Z^{2,1}$ as indicated.
    Survival curves are shown for four induction therapies.
    Black, red, green and blue
    curves indicate $Z^1=$ FAI, FAI+ATRA, FAI+GCSF and
    FAI+ATRA+GCSF, respectively. }} 
\label{fig:data1}
\end{figure}

We first report results for $T^{(R,D)}$. 
Of 210 patients, 39 (18.57\%) experienced resistance to their
induction therapies. 
The rate of resistance varied across regimes, from
31\% for patients receiving FAI, 24\% for FAI plus ATRA,
7.8\% for FAI plus GCSF,  and 10\% for FAI plus ATRA plus GCSF.
The times to treatment
resistance were longer, with greater variability in the FAI plus
GCSF arm compared to  the other three arms. Among the 39 patients who
were resistant to induction therapies, 27 were given HDAC as salvage
treatment,  of whom  2 were censored before observing death. 
Figure \ref{fig:data1} summarizes survival regression under the
proposed DDP-GP model by plotting posterior predicted survival
functions for a hypothetical future patient
at scaled age 0 with poor prognosis cytogenetic abnormality. 
The figure shows posterior predicted survival
functions, arranged by different induction therapies $Z^1$ 
(the four curves in each panel),
$T^{(0,R)}$ and $Z^{2,1}$ (as indicated in the subtitle). 
From Figure
\ref{fig:data1}, we can see that patients with shorter $T^{(0,R)}$ 
have lower predicted survival 
once their cancer became resistant. Also, patients with
$s_1=R$ who received $Z^{2,1}$  = HDAC as salvage had worse survival predication 
than patients who received salvage treatment with non HDAC. Similar results can be obtained for other transition times.
Inferences for similar survival regressions for $T^{(C,P)}$ and $T^{(P,D)}$ are
summarized in the on-line supplement.

\begin{table}
\centering
\begin{tabular}{ |l|c|c|l|l|c| }
\hline
&\multicolumn{2}{ |c| }{Resistance}&\multicolumn{3}{ |c| }{Die after resistance} \\
Induction&$N$&$T^R$(days)&Salvage&$N$&$T^{RD}$(days)\\ \hline
All &39&59 (47,84)&All&37&76 (27,187) \\
FAI &17&63 (41,97)&HDAC&25&65 (21,154)\\
FAI+ATRA &13&59 (55,76)&&& \\
FAI+GCSF&4&77 (43.5,106.75)&non HDAC &12&146 (79, 376.75) \\
FAI+ATRA+GCSF &5&51 (48, 65)&&&  \\
\hline
\end{tabular}
\vskip .2in
\begin{tabular}{ |l|c|c|l|l|l|l|l|c|c| }
\hline
&\multicolumn{2}{ |c| }{CR}&\multicolumn{3}{ |c| }{Die after progression}\\
Induction&$N$&$T^C$(days)&Salvage&$N$&$T^{PD}$(days)\\ \hline
All &102&32 (27,41)&All&83&120 (45,280)\\
FAI &20&31 (29, 44)&HDAC&47&106 (45,175.5)\\
FAI+ATRA &26&31 (25.25, 35)&&& \\
FAI+GCSF&28&35.5 (28,42.75)&non HDAC& 36&147.5 (42.75, 592.25) \\
FAI+ATRA+GCSF &28&32 (26,41)&&& \\
\hline
\end{tabular}
\caption{The sample median of each transition time is given, with
  lower 25\% quantile and upper 75\% quantile in the parenthesis next
  to each median . } 
\label{table:table1}
\end{table}

\subsection{Estimating the Regime Effects}
\begin{figure}[!h]
\centering
\includegraphics[width=\textwidth]{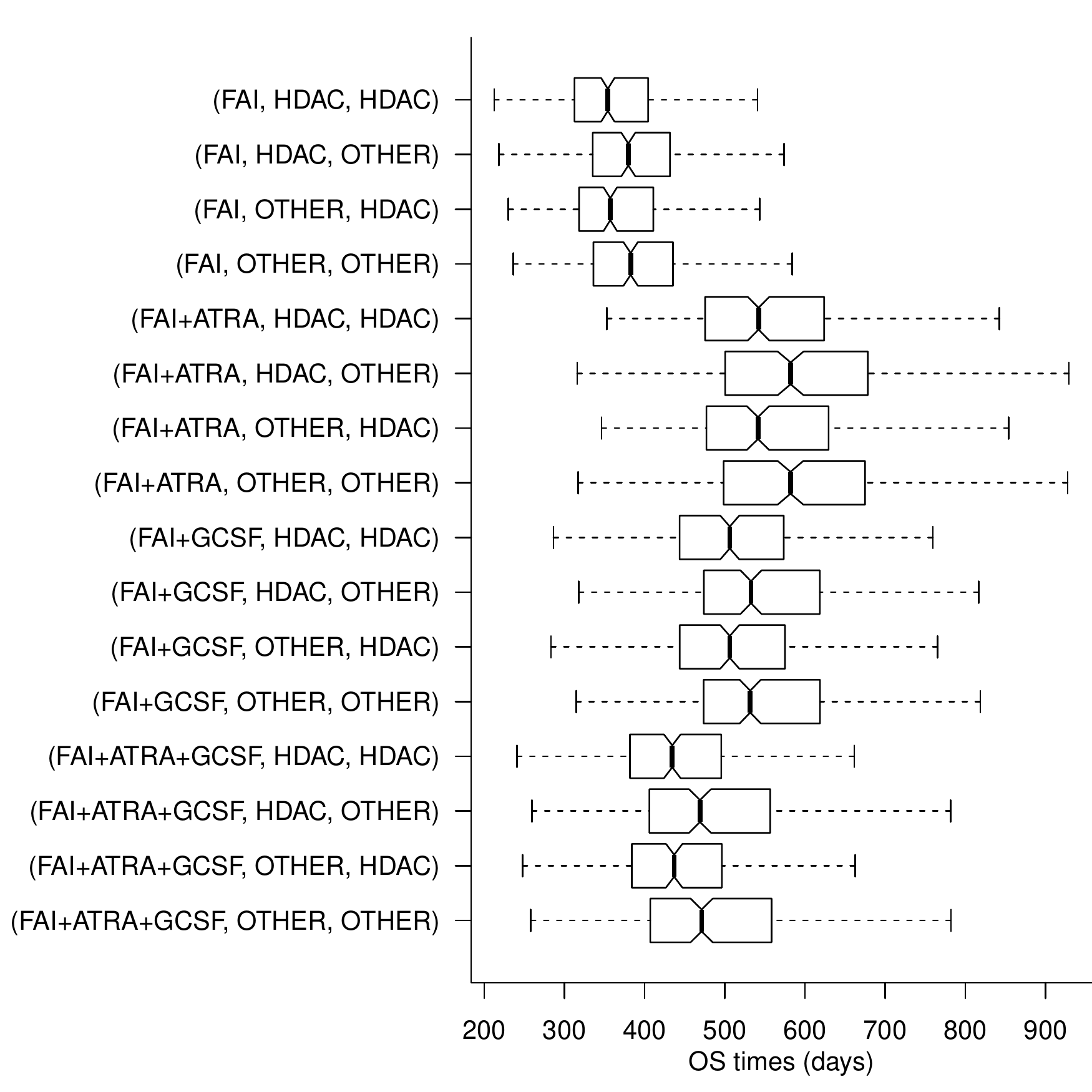}
\caption{Marginal posterior distributions of overall survival time
  under the DDP-GP model for all 16 regimes.} 
\label{fig:data4}
\end{figure}

In the AML-MDS trial, the four induction therapies and two salvage
therapies define a total 16 regimes. The mean survival time
estimates under each of the 16 regimes were calculated using 
posterior inference under independent DDP-GP models 
$F^k(Y_i^k \mid \bx_i^k)$ for each of
the $\ntrans=7$ transition times.
For comparison we also evaluated mean survival times under
the IPTW method. See equation (\ref{eq:IPTW}) in the Appendix for
details.  
Table \ref{table:table2} summarizes the results under IPTW and
under the DDP-GP model (including 90\% credible intervals).
Figure (\ref{fig:data4}) shows boxplots of the marginal posterior
distributions of survival times under the DDP-GP model for the same 16
regimes.  
\begin{table}
\begin{tabular}{ |l|l|c|c| }
\hline
Regime $(A, B_1, B_2)$&\multicolumn{3}{ |c| }{Estimated mean OS times (days)} \\
&\multicolumn{3}{ |c| }{DDP-GP} \\
&IPTW&Posterior mean& $90\%$ CI\\ \hline
(FAI, HDAC, HDAC) & 191.67&390.35&  (286.47 \  545.6)  \\
(FAI, HDAC, other) & 198.18&416.34& (295.84  \ 581.73) \\
(FAI, other, HDAC) & 216.59& 394.2&(287.15  \ 538.63)\\
(FAI, other, other) & 222.42&420.19&(296.51  \ 579.05)\\ \hline
(FAI+ATRA, HDAC, HDAC) & 527.43&572.9&(416.63  \ 829.12)  \\
(FAI+ATRA, HDAC, other) & 458.85&617.15&(434.4  \ 905.82)\\
(FAI+ATRA, other, HDAC) & 532.29& 573.46&(413.59  \ 830.39) \\
(FAI+ATRA, other, other) & 464.39& 617.71&(434.49  \ 900.32) \\ \hline
(FAI+GCSF, HDAC, HDAC) & 326.15&542.06&(393.49  \ 725.23) \\
(FAI+GCSF, HDAC, other) & 281.78&578.24&(419.69  \ 781.05) \\
(FAI+GCSF, other, HDAC) & 327.66&542.5&(392.77  \ 726.08)\\
(FAI+GCSF, other, other) & 283.36& 578.68&(421.46 \  781.26) \\ \hline
(FAI+ATRA+GCSF, HDAC, HDAC) & 337.44&458.34&(327.91  \ 651.21) \\
(FAI+ATRA+GCSF, HDAC, other) & 285.64&502.48&(360.29 \  727.44)\\
(FAI+ATRA+GCSF, other, HDAC) & 362.56&459.42&(328.09 \  651.61) \\
(FAI+ATRA+GCSF, other, other) & 309.62&503.56&(358.84  \ 726.88)\\ \hline
\end{tabular}
\caption{Mean overall survival time under the IPTW method and the
  posterior mean and 90\% credible interval (CI) under the DDP-GP model.} 
\label{table:table2}
\end{table}

The two methods give very different estimates for mean survival time,
with the DDP-GP likelihood-based estimator larger than the
corresponding IPTW estimator for most regimes. The differences are
expected because of the distinct properties of these two methods. The
IPTW estimator uses the covariates to estimate the regime probability
weights. In contrast, the DDP-GP likelihood-based method computes mean
survival time, using G-estimation,  accounting for patients'
covariates and previous transition times in addition to treatment
followed by marginalizing over the empirical covariate distribution to obtain
$\eta({\bf Z})$. Additionally, the IPTW estimate is calculated from
the overall samples, whereas the likelihood-based DDP-GP method models each
transition time distribution separately, which reduces the effective
sample size for each model fit and thus increases the overall
variability even though they share the same prior for the
$\bbeta^k$'s. 

Under both methods, the estimates were smallest for the four regimes
with FAI as induction therapy regardless of salvage treatment, and the
90\% credible intervals were relatively small for these inferior
regimes. Under the IPTW method, the estimates were largest for the
four regimes with FAI plus ATRA as induction therapy, and the best
regime is (FAI+ATRA, other, HDAC). With the DDP-GP likelihood-based
approach, FAI plus ATRA as induction also gave the largest estimates,
except for the regimes (FAI+GCSF, HDAC, other) and (FAI+GCSF, other,
other), while the best regime is (FAI+ATRA, other, other). Most
importantly, the DDP-GP likelihood-based approach showed that  (FAI +
ATRA, $Z^{2,1}$, other) was superior to (FAI + ATRA, $Z^{2,1}$,
HDAC) regardless of $Z^{2,1}$. 
Therefore, our results suggest that (1) FAI plus ATRA was the best
induction therapy, (2)  if the patient's disease was resistant to FAI
plus ATRA, then it was irrelevant whether the salvage therapy
contained HDAC, and (3)  if patients experienced progression after
achieving CR with FAI plus ATRA, then salvage therapy with non HDAC
was superior. 

These conclusions, although not confirmatory, are contradictory with
those given by   \cite{estey1999randomized}, who concluded that
none of the three adjuvant combinations FAI plus ATRA, FAI plus
GCSF, or FAI plus ATRA plus GCSF were significantly different from FAI
alone with respect to either survival or event-free survival time,
based on consideration of  only the front-line therapies by applying conventional Cox
regression and hypothesis testing.

\section{Conclusions}
\label{sec:dis}
We have proposed a Bayesian nonparametric DDP-GP model for analyzing survival data and evaluating joint effects of induction-salvage therapies in clinical trials, using  the posterior estimates, to predict survival for future patients. The Bayesian paradigm works very well, and the simulation studies suggest that our DDP-GP method yields more reliable estimates than IPTW.

We employed two different methods to evaluate the 16 possible two-stage regimes for choosing induction and salvage therapies in the leukemia trial data. The IPTW method estimates the regime effect by using  covariates only to compute the assignment probabilities of salvage therapies to correct for bias. In contrast, likelihood-based G-estimation under the DDP-GP model accounts for all possible outcome paths, the transition times between successive states, and effects of covariates and previous outcomes, on each transition time. Although the two methods gave different numerical estimates of mean survival time, they both reached the conclusion that FAI plus ATRA was the best induction therapy and FAI was the worst induction therapy. Although our current models are set up for two-stage treatment regimes, they easily can be extended to other applications with multi-stage regimes.

\section*{Acknowledgements}
This research was supported by NCI/NIH grant  R01 CA157458-01A1 (Yanxun Xu and Peter M\"uller) and R01 CA83932 (Peter F. Thall).

\section*{Appendix}

The following structure is that given by \cite{wahed2013evaluating}, and is included here for completeness.
The risk sets of the seven transition time in the leukemia trial are defined as follows.
Let $G^0=\{1,\ldots,n\}$ denote the initial risk set at the start of induction chemotherapy,
and $G^{(0,r)} =\{i:\; s_{1i}=r\}$ for $r=D, C, R$, so  $G^0$ = $G^{(0,D)}\cup G^{(0,C)}\cup G^{(0,R)}$.
Similarly,  $G^{(C,P)}=\{i:\; s_{1i}=C, s_{2i}=P\}$ is the later risk set for $T^{(P,D)}$.

To record right censoring, let $U_i$ denote the time from the start of
induction to last followup for patient $i$.  We assume that
$U_i$  is conditionally independent of the transition times given prior
transition times and other covariates.
 Censoring of event times occurs by competing risk and/or loss to
follow up. For a patient $i$ in the risk set for event time $T^k_i$,
let $\delta_i^k$ = 1 if a patient $i$ is not
censored and 0 if patient $i$ is right censored.
For example, $\delta_i^{(0,D)} = 1$ for $i \in G^0$ if $T_i^{(0,D)} = \min(U_i,
T_i^{(0,k)}, k=D,C,R)$.
Similarly, $\delta_i^{(R,D)}=1$ for $i \in G^{(0,R)}$ if $T_i^{(0,R)}+T_i^{(R,D)} <
U_i$ and $\delta_i^{(P,D)}=1$ for $i \in G^{(C,P)}$ if
$T_i^{(0,C)}+T_i^{(C,P)}+T_i^{(P,D)} < U_i$.

Let $V_{x,i}$ denote the observed time for
patient $i$ in risk set $G^x$, as follows.
For $i \in G^0$ let $V_{1,i}=\min(T^{(0,D)}_i,T^{(0,R)}_i,T^{(0,C)}_i,U_i)$ denote the
observed time for the stage 1 event or censoring.
For $i \in G^{(0,C)}$ let $V_{Ci}=\min(T^{(C,D)}_i,T^{(C,P)}_i, U_i-T^{(0,C)}_i)$ denote
the observed event time for the competing risks $D$ and $P$
and loss to followup.
Similarly, for $i \in G^{(0,R)}$, let $V_{Ri} = \min(T^{(R,D)}_i, U_i-T^{(R,D)}_i)$, and
for $i \in G^{(C,P)}$ let $V_{(C,P),i}=\min(T^{(P,D)}_i, U_i-T^{(0,C)}_i-T^{(C,P)}_i)$.

The joint likelihood function is the product
$
  {\cal L}\ =\ {\cal L}_1{\cal L}_{2}{\cal L}_{3}{\cal L}_{4}.
$
The first factor ${\cal L}_1$ corresponds to response to induction therapy,
\begin{equation}
  {\cal L}_1=\prod_{i \in G^0}
  \prod_{k\in \{D, R, C\}}
  f^{(0,k)}( V_{1,i} \mid \bx^k_i)^{\delta^{(0,k)}_i}
  S^{(0,k)}(V_{1,i} \mid \bx^k_i)^{1-\delta^{(0,k)}_i}.
  \label{eq:like1}
\end{equation}
where $S^k$ = $1-F^k.$
The second factor ${\cal L}_2$ corresponds to patients $i \in G^{(0,R)}$ who
experience resistance to induction and receive salvage $Z^{2,1}$,
\begin{equation}
  {\cal L}_{2}=\prod_{i\in G^{(0,R)}}
  f^{(R,D)}(V_{Ri} \mid \bx^{(R,D)}_i)^{\delta^{(R,D)}_i}
  S^{(R,D)}(V_{Ri} \mid \bx_i^{(R,D)})^{1-\delta^{(R,D)}_i}.
\label{eq:like2}
\end{equation}
The third factor ${\cal L}_3$  is the
likelihood contribution from patients achieving CR,
\begin{equation}
  {\cal L}_{3} = \prod_{i\in G^{(0,C)}}\;\; \prod_{k=(C,D),(C,P)}
  f^{k }(V_{Ci} \mid \bx^{k}_i)^{\delta^{k}_i}
  S^{k}(V_{Ci} \mid \bx^{k}_i)^{1-\delta^{k}_i} .
\label{eq:like3}
\end{equation}
The fourth factor ${\cal L}_4$ is the contribution from
patients who experience tumor progression after CR
\begin{equation}
  {\cal L}_{4}=
  \prod_{i\in G^{(C,P)}}
  f^{(P,D)}(V_{CP,i} \mid \bx^{(P,D)}_i)^{\delta^{(P,D)}_i}
  \,
  S^{(P,D)}(V_{CP,i} \mid \bx^{(P,D)}_i)^{1-\delta^{(P,D)}_i}.
\label{eq:like4}
\end{equation}

The mean survival time of a patient treated with regime ${\bf Z}$  = $(Z^1, Z^{2,1}, Z^{2,2})$
is
\begin{multline}
\eta({\bf Z})
  =\int \Big{[} p(s_{1}=D \mid \bx^0, Z^1)\eta^{(0,D)}(\bx^0, Z^1)\Big{]}d\phat(\bx^0) \\
  +  \int \Big\{p(s_{1}=R \mid  \bx^0, Z^1)\Big{[}\eta^R( \bx^0, Z^1)+
     \int       \eta^{(R,D)}(\bx^0, Z^1, Z^{2,1}, T^{(0,R)})d\mu(T^{(0,R)})\Big{]} \Big\}d\phat(\bx^0) \\
 +  \int p(s_{1}=C \mid \bx^0, Z^1) \Big{[}\eta^C( \bx^0, Z^1)+\int \Big{[}
     p(s_2=D\mid s_1=C, \bx^0, Z^1, T^{(0,C)})\eta^{(C,D)}( \bx^0, Z^1, T^C)    \\
   +  p(s_2=P\mid s_1=C, \bx^0, Z^1, T^{(0,C)})[\eta^{(C,P)}( \bx^0, Z^1, T^{(0,C)})    \\
   +           \int \eta^{(P,D)}( \bx^0, Z^1, Z^{2,2}T^{(0,C)}, T^{(C,P)})d\mu(T^{(C,P)})]
               d\mu(T^{(0,C)})\Big{]}d\phat(\bx^0).
\label{eq:likebased}
\end{multline}

We compute
the IPTW estimates for overall mean survival with regime  ${\bf Z}$ as
\begin{equation}
IPTW({\bf Z})=\sum_{i=1}^n\, w_i ({\bf Z}) T_i\ /\ \sum_{i=1}^n\, w_i ({\bf Z}), 
\end{equation}
where
\beq
w_i ({\bf Z})  &=&   \frac{I({\bf Z}={\bf Z}_i)
\delta_i}{\hat{K}(U_i)}\biggl[I(s_{1i}=D)+I(s_{1i}=R)I_i(Z^{2,1})/\hat{\mathrm{Pr}}(Z^{2,1}\mid s_{1i}=R, Z^1, \bx^0_i, T_i^{(0,R)})\nonumber\\
&&+I(s_{1i}=C, s_{2i}=D)\nonumber\\
&&+I(s_{1i}=C, s_{2i}=P)I_i(Z^{2,2})/\hat{\mathrm{Pr}}(Z^{2,2}\mid s_{1i}=C, s_{2i}=P, Z^1, \bx_i^0, T_i^{(0,C)}, T_i^{(C,P)})\biggr]. \nonumber\\
\label{eq:IPTW}
\eeq
In (\ref{eq:IPTW}), $\hat{K}$ is the Kaplan-Meier estimator of the
censoring survival distribution $K(u)=P(U\geq t)$ at time
$t$. $I_i(Z)$ is is an indictor of treatment $Z$ and 0 otherwise, and  $\hat{\mathrm{Pr}}(Z^{2,1}\mid s_{1i}=C, Z^1, \bx^0_i, T_i^{(0,R)})$ is the
probability of receiving salvage treatment $Z^{2,1}$ estimated using logistic
regression, and similarly for $\hat{\mathrm{Pr}}(Z^{2,2}\mid s_{1i}=C,
s_{2i}=P, Z^1, \bx^0_i, T_i^{(0,C)}, T_i^{(C,P)})$. The above estimator has been
shown to be consistent under suitable assumptions
\citep{wahed2013evaluating, scharfstein1999adjusting}.

\bibliographystyle{apalike}
\bibliography{causal-DDP,Gaussian}
\clearpage

\end{document}